\documentclass[12pt,fleqn]{article}

\usepackage{amsmath,amssymb,amsfonts,graphicx,epsfig}
\def\bea{\begin{eqnarray}}
\def\eea{\end{eqnarray}}
\def\be{\begin{equation}}
\def\ee{\end{equation}}
\def\ba{\begin{array}}
\def\ea{\end{array}}
\def\nn{\nonumber}
\def \lsim{\mathrel{\vcenter
{\hbox{$<$}\nointerlineskip\hbox{$\sim$}}}}
\def \gsim{\mathrel{\vcenter
{\hbox{$>$}\nointerlineskip\hbox{$\sim$}}}}

\DeclareFontFamily{OT1}{pzc}{}
\DeclareFontShape{OT1}{pzc}{m}{it}{<-> s * [1.4] pzcmi7t}{}
\DeclareMathAlphabet{\mathpzc}{OT1}{pzc}{m}{it}
\DeclareMathAlphabet{\mathscr}{OT1}{pzc}{m}{it}

\begin{document}
\pagestyle{plain}

\renewcommand{\theequation}{\arabic{section}.\arabic{equation}}
\setcounter{page}{1}

\begin{titlepage} 

\rightline{\footnotesize{CERN-PH-TH/2012-238}} \vspace{-0.2cm}

\begin{center}

\vskip 0.4 cm

\begin{center}
{\LARGE{ \bf Metastable de Sitter vacua in N=2 \\[3mm] to N=1 truncated supergravity}}
\end{center}

\vskip 1cm

{\large 
Francesca Catino$^{a,b}$, Claudio A. Scrucca$^{a}$ and Paul Smyth$^{a}$
}

\vskip 0.5cm

{\it
$^{a}$Institut de Th\'eorie des Ph\'enom\`enes Physiques, EPFL, \\
\mbox{CH-1015 Lausanne, Switzerland}\\
$^{b}$Theory Division, Physics Department, CERN, \\
CH-1211 Geneva 23, Switzerland\\
}

\vskip 1.5cm

\end{center}

\begin{abstract}

We study the possibility of achieving metastable de Sitter vacua in general N=2 to N=1 truncated 
supergravities without vector multiplets, and compare with the situations arising in N=2 theories with 
only hypermultiplets and N=1 theories with only chiral multiplets. In N=2 theories based 
on a quaternionic manifold and a graviphoton gauging, de Sitter vacua are necessarily 
unstable, as a result of the peculiar properties of the geometry. In N=1 theories based on a 
K\"ahler manifold and a superpotential, de Sitter vacua can instead be metastable provided the geometry 
satisfies some constraint and the superpotential can be freely adjusted. In N=2 to N=1 truncations, 
the crucial requirement is then that the tachyon of the mother theory be projected out from the 
daughter theory, so that the original unstable vacuum is projected to a metastable vacuum. 
We study the circumstances under which this may happen and derive general constraints for metastability
on the geometry and the gauging. We then study in full detail the simplest case of quaternionic 
manifolds of dimension four with at least one isometry, for which there exists a general parametrization, 
and study two types of truncations defining K\"ahler submanifolds of dimension two. As an application, 
we finally discuss the case of the universal hypermultiplet of N=2 superstrings and its truncations to the 
dilaton chiral multiplet of N=1 superstrings. We argue that de Sitter vacua in such theories are necessarily 
unstable in weakly coupled situations, while they can in principle be metastable in strongly coupled regimes. 

\end{abstract}

\bigskip

\end{titlepage}

\newpage

\section{Introduction} \setcounter{equation}{0}

Realizing a de Sitter vacuum which is at least metastable and where supersymmetry is spontaneously broken, 
as required by particle phenomenology and cosmological observation, has proven to be very difficult within the 
context of string theory, and no completely convincing setup realizing 
such a vacuum has been singled out so far. It is however by now well understood that a major source of obstruction 
to metastability stems just from the peculiar general structure of supergravity, which represents the general 
framework for a low-energy effective description of string theory. To make further progress in discriminating string 
models, it would then be highly desirable to have a complete understanding of the restrictions arising within supergravity 
on the possibility of achieving such a viable vacuum. Ideally, this should moreover encompass not only the case of 
minimal supersymmetry, which is directly interesting for model building, but also the various cases of extended 
supersymmetry, which partly reflect  some of the additional special features displayed by models with 
a higher-dimensional origin.

The analysis of the conditions under which de Sitter vacua may be at least metastable in supergravity is complicated 
by the fact that the general form of the mass matrix for the scalar fluctuations around such a vacuum depends in a rather 
intricate way on the free parameters in the Lagrangian. Studying the full mass matrix and translating the requirement of its 
positivity into necessary and sufficient constraints on these parameters is then unfortunately possible only on a model by 
model basis. However, one may specialize this kind of analysis to a restricted set of particularly dangerous modes, and 
try in this way to derive some constraints that are only necessary and not sufficient but more general and useful. 
More precisely, one may try to exploit the restrictions on the structure of the kinetic energy, and ignore instead the 
details of the source of potential energy. To do so, one should focus on those scalar modes for which the mass 
happens to be constrained independently of the precise form of the potential, as a result 
of the assumed spontaneous breaking of supersymmetry and possibly also of some of the internal symmetries.
The a priori most dangerous modes are those belonging to multiplets for which a supersymmetric mass term 
is either forbidden or strongly constrained. These include all the real scalar partners of the would-be Goldstino fermions 
associated to broken supersymmetry \cite{DD,GRS1,GRS2}, as well as those of the would-be Goldstone bosons associated to 
broken internal symmetries \cite{GRS3}, which we shall simply call sGoldstini and sGoldstones. Indeed, such modes have 
masses that are controlled by supersymmetry or internal symmetry breaking effects, whose form is very constrained.

The situation in N=1 supergravity is well understood. For theories with only chiral multiplets, based on a K\"ahler manifold 
and an arbitrary superpotential, the only dangerous modes are the two sGoldstini. The positivity of their average 
mass implies a simple and sharp necessary condition on the K\"ahler manifold: this should admit points and directions for 
which the sectional curvature is larger than a certain critical value depending on the vacuum energy and the gravitino mass 
\cite{GRS1,GRS2} (see also \cite{CGRGLPS1,CGRGLPS2}). For theories with chiral and vector multiplets, based on a K\"ahler 
manifold with isometries gauged by vector bosons and a gauge invariant superpotential, the dangerous modes are not only the 
two sGoldstini but also all the sGoldstones. The positivity of the average mass of the two sGoldstini implies again a condition on 
the K\"ahler manifold: this should admit points and directions for which the sectional curvature is larger than a certain critical value 
depending now also on the gauging data \cite{GRS3}. On the other hand, the positivity of the masses of the sGoldstones does 
not seem to imply any simple and general constraint, although examples are known where they can lead to instabilities on their own. 
The situation in the rigid limit is also well understood, and has been discussed in \cite{JS} and also \cite{BS}, where it was argued 
that the lightest scalar in the theory is in general a combination of sGoldstini and sGoldstones.

The case of N=2 supergravity is more complex and only partly understood. For theories with only hypermultiplets, based 
on a quaternionic manifold with an isometry gauged by the graviphoton, the only dangerous modes are the four 
sGoldstini, out of which one is absorbed by the graviphoton in a Higgs mechanism giving it mass and only three 
represent physical scalar modes. Their average mass turns out to be given by a completely universal value, 
which depends only on the vacuum energy and the gravitino mass and turns out to be negative, and as a result metastability 
is impossible to achieve: there is always at least one sGoldstino that has a negative square mass \cite{GRLS}. In theories with 
only vector multiplets and Abelian gaugings, based on a special-K\"ahler manifold with commuting isometries gauged by vector 
bosons, the dangerous modes are only the two sGoldstini, since there is no way the Abelian symmetries can be broken. 
Their average mass turns again out to be given by a completely universal value, which depends as before only on the vacuum 
energy and the gravitino mass and turns out to be negative, and as a result metastability is again impossible to achieve: there 
is always at least one sGoldstino with a negative square mass \cite{Many}. For more general  theories involving hyper and 
vector multiplets and/or non-Abelian gaugings, the situation is more complicated and the dangerous modes are 
in principle not only the sGoldstini but also all the sGoldstones. No simple general necessary condition for metastability has 
been derived so far for this case, but a few particular examples of models admitting metastable de Sitter vacua are known 
\cite{FTV,O}, and a non-trivial criterion is thus expected to emerge. The situation in the rigid limit is slightly better but still only 
partly understood. In this limit, it has been possible to rederive the above no-go theorems in a simpler way and 
generalize the computation of the average mass of the sGoldstini to the case of theories with only vector multiplets but generic 
non-Abelian gaugings, showing that in that case it can be positive if there are Fayet-Iliopoulos terms for Abelian factors \cite{JS}. 
Moreover, it has been argued in \cite{AB} that the N=2 supersymmetry algebra does not allow for a consistent non-linear 
realization whenever the theory possesses an $SU(2)_R$ symmetry, that is whenever Fayet-Iliopoulos terms are absent.

The cases of N=4 and N=8 supergravities are qualitatively different and again only partly understood. The main novelty in these 
cases is that the scalar manifold is completely fixed to be a definite coset space, and the only freedom one has is to gauge 
suitable isometries with the graviphotons and/or vector bosons. The dangerous modes are a priori the various sGoldstini
and all the sGoldstone. A systematic analysis of the average sGoldstino mass was performed respectively in \cite{BR} and 
\cite{BLR}, and showed that its positivity puts quite strong constraints on the gauging. No similar study was performed 
so far for the sGoldstone masses. On the other hand, all the known examples of de Sitter vacua in such theories happen to be 
unstable (see for instance \cite{DWP,DWPT} and \cite{HW,KLPS,DI}) and it is conceivable that one might eventually be able 
to prove a no-go theorem. The situation in the rigid limit is in this case totally trivial. The N=4 theory requires a flat scalar 
manifold and its potential does not allow supersymmetry breaking critical points. The N=8 theory, on the other hand, has a 
totally empty rigid limit.

The aim of this paper is to initiate a similar systematic study of the conditions for the metastability of de Sitter vacua in supergravities 
where the amount of supersymmetry is reduced through a consistent truncation. This situation is in fact directly realized in several 
string constructions, and may thus provide a more realistic and representative framework to study the situation for string-derived models. 
The crucial new aspect arising in such a context is that one may start from an unstable de Sitter vacuum of the mother theory and obtain 
a metastable de Sitter vacuum in the daughter theory, provided the original tachyon is projected out by the truncation. This possibility was 
explored in \cite{RR} for the case of N=8 to N=4 truncations, where the instability was found to always persist, and for the case of 
N=4 to N=2 truncations, where the instability was shown to disappear in some particular cases. Here we would like to study in some 
generality the basic case of N=2 to N=1 truncations. For concreteness, we shall restrict to N=2 theories with only hypermultiplets 
truncated to N=1 theories which then involve only chiral multiplets. In this simplest situation, we already know that any de Sitter vacuum 
of the mother N=2 theory is necessarily unstable and involves at least one tachyon, while de Sitter vacua in the N=1 daughter theory 
can be metastable and free of tachyons. Moreover, the sGoldstino masses and the consistency conditions for the truncation are controlled 
by purely geometric quantities, and it should therefore be possible to characterize in simple and general terms the possibility of starting 
with a vacuum possessing a single tachyonic sGoldstino and projecting out this mode through a truncation. 

The paper is organized as follows. In sections 2 and 3 we briefly review the metastability conditions emerging in N=1 theories with 
only chiral multiplets and N=2 theories with only hypermultiplets. In section 4 we derive the metastability conditions in 
truncations of N=2 theories with only hypermultiplets to N=1 theories with only chiral multiplets. In section 5 we study in detail the 
simplest case of N=2 theories with a single hypermultiplet, based on a generic quaternionic space with at least one isometry,
for which there exists a general explicit description as a Przanowski-Tod space, and describe two different kinds of truncations to 
N=1 theories with a single chiral multiplet. In section 6 we illustrate our results with a few specific examples of this type, which 
are directly relevant for the low-energy effective description of the universal hypermultiplet of string models. In section 7 we 
present our conclusions. In appendix A, we summarize the relevant details about the geometry of Przanowski-Tod spaces.

\section{N=1 supergravity with chiral multiplets}\setcounter{equation}{0}

In N=1 supergravity with $n$ chiral multiplets, the geometry of the scalar manifold that controls the 
kinetic terms through its metric $g_{i \bar \jmath}$ is restricted to be K\"ahler, and the source 
of potential $V$ is represented by an arbitrary holomorphic superpotential. In units where $\kappa =1$,
the part of the Lagrangian describing the $n$ complex scalar fields $\phi^i$ takes the following form:
\bea
{\cal L} = - g_{i \bar \jmath}\, \partial^\mu\hspace{-1pt} \phi^i \partial_\mu \phi^{\bar \jmath} - V \,.
\eea

Let us first review the properties of the geometry. The holonomy group is $U(n)$. 
The vielbein is written as $e_i^{a}$ and its conjugate is $e_{\bar \imath}^{\bar a} = (e_i^a)^*$.
The indices $i,\bar \imath=1,\cdots,n$ refer to the manifold while the indices  
$a,\bar a=1,\cdots,n$ refer to the tangent space. The tangent space metric is simply 
$g_{a \bar b} = \delta_{a \bar b}$. The manifold metric is then given by:
\bea
g_{i \bar \jmath} = \delta_{a \bar b} e_i^{a} e_{\bar \jmath}^{\bar b} \,.
\eea
There is one complex structure satisfying $J^i_{\;\,k} J^k_{\;\,j} = - \delta^i_j$. It is given by
\bea
J_{i \bar \jmath} = i \delta_{a \bar b} e_i^{a} e_{\bar \jmath}^{\bar b} = i g_{i \bar \jmath} \,.
\eea
The $U(n)$ connection $\Gamma_{i}$ is determined by the torsion-free constraint on the vielbeins, 
which reads $\nabla_i e_j^{a} = 0$ in terms of a covariant derivative including both this connection 
and the Christoffel one. Its curvature two-form can be parametrized in terms of a tensor $R_{a \bar b c \bar d}$ 
which has to be symmetric in its holomorphic and antiholomorphic indices but is otherwise arbitrary:
$R^{a \bar b}_{i \bar \jmath} = e_i^c e_{\bar \jmath}^{\bar d} R^{a \bar b}{\!}_{c \bar d}$.
The Riemann curvature tensor with flat indices is then simply $R_{a \bar b c \bar d}$, while its 
version with only curved indices is instead given
\be
R_{i \bar \jmath p \bar q} = e_i^{a} e_{\bar \jmath}^{\bar b} e_p^{c} e_{\bar q}^{\bar d} 
R_{a \bar b c \bar d} \,.
\ee
The Ricci and scalar curvatures are given by the contractions $R_{i \bar \jmath} = - g^{p \bar q} R_{i \bar \jmath p \bar q}$
and $R_{\rm sca} = g^{i \bar \jmath} R_{i \bar \jmath}$.
The Weyl curvature is instead controlled by the traceless part of $R_{i \bar \jmath p \bar q}$. Finally, one has
the standard Ricci decomposition
\bea
R_{i \bar \jmath p \bar q} =
\frac {2}{(n+1)(n+2)} g_{i (\bar \jmath} \hspace{2pt} g_{p \bar q)} R_{\rm sca}
- \frac {2}{n+2} \big(g_{i (\bar \jmath} \hspace{1pt} R_{p \bar q)} + g_{p (\bar q} \hspace{1pt} R_{i \bar \jmath)} \big) 
+ C_{i \bar \jmath p \bar q}\,.
\eea

Let us next describe the properties of the superpotential $W$ that is used to generate a potential. 
The only restriction is that it should be holomorphic. It is 
then convenient to introduce the quantities 
\bea
L = e^{K/2} W \,,\;\; N_i = e^{K/2} \big(W_i + K_i W \big) \,.
\eea
These satisfy
\bea
\nabla_{\bar \jmath} L = 0 \,,\;\; \nabla_i L = N_i \,.
\eea
Moreover:
\bea
\big[\nabla_i, \nabla_{\bar \jmath}\big] L = - g_{i \bar \jmath} L \,,\;\;
\big[\nabla_i,\nabla_{\bar \jmath}\big] N_p = R_{i \bar \jmath p \bar q} \bar N^{\bar q} - g_{i \bar \jmath} N_p \,.
\label{commN1}
\eea
The scalar potential then takes the following form:
\bea
V = \bar N^i N_i - 3 |L|^2 \,.
\eea
Under supersymmetry transformations, the gravitino and the chiralini transform as 
$\delta \psi_\mu = i L \gamma_\mu \epsilon + \cdots$ and $\delta \chi^a = N^a \epsilon + \cdots$,
where $N_a = U_a^i N_i$. The supersymmetry breaking scale is thus the norm of 
$N_i$, $M_{\rm susy} = |N|$, while the gravitino mass is given by the norm of $L$,
$m_{3/2} = |L|$. The Goldstino is $\chi = N_a \chi^a$, while the complex sGoldstino is 
$\phi = N_i \phi^i$. 

The mass matrix of the scalars at a stationary point where $\nabla_i V = 0$ is given by 
$m^2_{i \bar \jmath} = \nabla_i \nabla_{\bar \jmath} V$ and $m^2_{ij} = \nabla_i \nabla_j V$. 
This is not the physical mass matrix, but the only additional thing to take into account is the 
non-trivial metric $g_{i \bar \jmath}$. The physical masses in the subspace defined by the 
complex direction $n^i = \bar N^i/|N|$ are given by $m^2_\pm = m^2_{i \bar \jmath} n^i n^{\bar \jmath} 
\pm |m^2_{i j} n^i n^j|$. A straightforward computation gives
\bea
m^2_\pm = R\, \bar N^k N_k + 2 |L|^2 \pm \big|\Delta\, \bar N^k N_k - 2\, \bar L^2\big|\,,
\eea
where
\bea
R = - \frac {R_{i \bar \jmath p \bar q} \bar N^i N^{\bar \jmath} \bar N^p N^{\bar q}}{(\bar N^k N_k)^2} \,,\;\;
\Delta = \frac {\nabla_i \nabla_j \nabla_k L\, \bar N^i \bar N^j \bar N^k}{(\bar N^w N_w)^2} \,.
\eea
This shows that the average of the masses of the two real sGoldstini is controlled by the sectional curvature 
$R$ of the scalar manifold in the plane defined by the two conjugate vectors $(N,\bar N)$, while their 
splitting is controlled by the quantity $\Delta$ depending on the third derivative of the superpotential along 
the direction $N$.

In terms of the scalar potential $V$ and the gravitino mass $m_{3/2}$, one finally finds that there average
sGoldstino mass $m^2_{\rm avr} = \frac 12 (m^2_+ + m^2_-)$ is simply given by:
\bea
m^2_{\rm avr} = R\,V + \big(3 R + 2\big)\, m_{3/2}^2 \,.
\eea
This quantity defines by construction an upper bound on the mass  squared of the lightest scalar. In order to have 
a metastable supersymmetry breaking vacuum with $V>0$, one then needs the sectional curvature $R$ to satisfy 
the following bound \cite{GRS1,GRS2,CGRGLPS1}:
\be
R > - \frac {2}{3 + V/m_{3/2}^2} > - \frac 23\,.
\ee
This represents a necessary condition for the existence of metastable de Sitter vacua on the choice 
of K\"ahler manifold. Indeed, the scalar manifold needs to admit at least one special point and one 
complex direction along which the sectional curvature is larger than the critical value $-\frac 23$. 
One can then easily show that once this condition is satisfied, it is always possible to arrange for the two 
sGoldstino masses to be degenerate and equal to $m^2_{\rm avr}$ while all the other states have larger square 
masses, by suitably tuning the superpotential. This means that for a given K\"ahler 
manifold the necessary and sufficient condition for the existence of metastable de Sitter vacua 
for some choice of the covariantly holomorphic section is that there exist a point and a complex direction 
such that $R > - \frac 23$.

\section{N=2 supergravity with hypermultiplets}\setcounter{equation}{0}

In N=2 supergravity with $n$ hypermultiplets, the geometry of the scalar manifold that controls the 
kinetic terms through its metric ${\mathpzc g}_{IJ}$ is restricted to be quaternionic \cite{BWN2}, and 
the only possible source of potential is the gauging of a triholomorphic isometry through the graviphoton. 
We follow \cite{ABCDFFM,DF}, and introduce an arbitrary real and negative parameter $\lambda$ 
in terms of which the negative scalar curvature of the scalar manifold is parametrized as 
${\cal R}_{\rm sca} = 8 n (n+2) \lambda$. 
\footnote{The relation to the parameter $\nu$ of  \cite{ACDV} is $\nu = 2 \lambda$.} 
In units where $\kappa =1$, the part of the Lagrangian describing the $4n$ real scalar fields $q^I$ and 
the graviphoton ${\cal A}_\mu$ then takes the following form:
\bea
{\cal L} = - \frac 18 {\cal F}_{\mu \nu} {\cal F}^{\mu \nu} + \lambda\, {\mathpzc g}_{IJ} D^\mu q^I D_{\mu} q^J - {\cal V} \,.
\eea
The covariant derivative is defined as $D_\mu q^I \equiv \partial_\mu q^I + {\cal A}_\mu k^I(q)$, where $k^I$ 
is the Killing vector generating the isometry, and the field strength reads 
${\cal F}_{\mu \nu} = \partial_\mu {\cal A}_\nu - \partial_\nu {\cal A}_\mu$.

Let us first recall the main properties of the geometry. The holonomy group is 
$SU(2) \times SP(2n)$. The vielbein can be written as ${\cal U}_I^{A \alpha}$ and satisfies the reality condition
$({\cal U}_I^{A\alpha})^* = {\cal U}_{I A \alpha} $. The index $I=1,\cdots,4n$ refers to the manifold while the indices
$A=1,2$ and $\alpha=1,\cdots,2n$ refer to the tangent space. The tangent space metric is given by 
${\mathpzc g}_{A\alpha B \beta} = \epsilon_{AB} c_{\alpha \beta}$, where $\epsilon_{AB}$ and $c_{\alpha \beta}$ 
denote the usual Levi-Civita and symplectic tensors, which are used in the standard way 
to raise and lower each type of flat sub-index. The manifold metric is then given by:
\bea
{\mathpzc g}_{IJ} = \epsilon_{AB} c_{\alpha \beta} \hspace{1pt} {\cal U}_I^{A\alpha} {\cal U}_J^{B \beta} \,.
\eea
There are three complex structures ${\cal J}^{x\hspace{1pt}I}_{\;\;\;\;\,J}$, with $x=1,2,3$, satisfying the algebra
${\cal J}^{x\hspace{1pt}I}_{\;\;\;\;\,K} {\cal J}^{y\hspace{1pt}K}_{\;\;\;\;\,J} = - \delta^{xy} \delta^I_J + \epsilon^{xyz} {\cal J}^{z\hspace{1pt}I}_{\;\;\;\;\,J}$. 
Denoting the usual Pauli matrices by $\sigma_A^{xB}$, they are given by:
\bea
{\cal J}^x_{IJ} = i \sigma^x_{AB} c_{\alpha \beta} \hspace{1pt} {\cal U}_I^{A \alpha} {\cal U}_J^{B \beta} \,.
\eea
The $SU(2) \times SP(2n)$ connection can be decomposed as 
$\Gamma^{A \alpha}_{I\hspace{1pt}B \beta} = \omega^A_{I\hspace{1pt}B} \delta^\alpha_\beta 
+ \Delta^\alpha_{I\,\beta} \delta^A_{B}$, where $\omega^A_{I\hspace{1pt}B} = \frac i2 \sigma^{xA}_{\;\;\;\;B}\, \omega_I^x$ 
and $\Delta^\alpha_{I\hspace{1pt}\beta}$ correspond to $SU(2)$ and $SP(2n)$ parts. It is determined by the torsion-free 
constraint on the vielbeins, which reads $\nabla_I {\cal U}_J^{A\alpha} = 0$ in terms of a covariant derivative including both this 
connection and the Christoffel one. The curvature two-form correspondingly reads 
${\cal R}^{A \alpha}_{IJ\hspace{1pt}B \beta} = K^{A}_{IJ\hspace{1pt}B} \delta^\alpha_\beta 
+ \Sigma^{\alpha}_{IJ\hspace{1pt}\beta} \delta^A_B$, where 
$K^{A}_{IJ\hspace{1pt}B} = \frac i2 \sigma^{xA}_{\;\;\;\;B}\, K^x_{IJ}$ 
and $\Sigma^\alpha_{IJ\hspace{1pt}\beta}$ correspond to $SU(2)$ and $SP(2n)$ parts. The general form 
that these are allowed to take can be parametrized in terms of a completely symmetric but otherwise arbitrary 
tensor ${\cal W}_{\alpha \beta \gamma \delta}$: $K^{AB}_{IJ} = i\lambda\, \sigma^{xAB}\! {\cal J}^x_{IJ} $ and 
$\Sigma^{\alpha \beta}_{IJ} = \epsilon_{CD}\, {\cal U}^{C \gamma}{\hspace{-13pt}}_{[I \raisebox{7pt}{$$}\hspace{5pt}} {\cal U}_{J]}^{D \delta} 
(2\lambda\, \delta^\alpha{\!\!\!}_{(\gamma} \delta^\beta{\!\!}_{\delta)} + {\cal W}^{\alpha \beta}{\!}_{\gamma\delta})$.
\footnote{Our curvature is defined in the standard way and corresponds to twice that of \cite{ABCDFFM,DF}. 
The relation with the symbols used in \cite{ABCDFFM,DF} is $K^x_{IJ} = 2\hspace{1pt} \Omega^x_{IJ}$, 
$\Sigma^{\alpha}_{IJ\hspace{1pt}\beta} = 2 {\mathbb R}^{\alpha}_{IJ\hspace{1pt}\beta}$, 
${\cal W}_{\alpha \beta \gamma \delta} = 2\hspace{1pt} \Omega_{\alpha \beta \gamma \delta}$. 
Moreover, we use the standard conventions for the normalization of differential forms.}
It follows that the Riemann curvature tensor with flat indices takes the form 
\bea
{\cal R}_{A\alpha B\beta C\gamma D\delta} 
= 2\lambda \big(\epsilon_{AB} \epsilon_{CD} c_{\alpha(\gamma} c_{\beta \delta)} 
+ \epsilon_{A(C} \epsilon_{BD)} c_{\alpha \beta} c_{\gamma \delta}\big)
+ \epsilon_{AB} \epsilon_{CD} {\cal W}_{\alpha \beta \gamma\delta} \,.
\eea
Its version with only curved indices is instead given by:
\be
{\cal R}_{IJPQ} = \lambda \big({\mathpzc g}_{I [P} {\mathpzc g}_{JQ]} + {\cal J}^x_{IJ} {\cal J}^x_{PQ} + {\cal J}^x_{I[P} {\cal J}^x_{JQ]}\big) 
+ {\cal W}_{IJPQ} \,,
\label{RiemannN2}
\ee
where
\be
{\cal W}_{IJPQ} = \epsilon_{AB} \epsilon_{CD}\, {\cal U}_I^{A\alpha} {\cal U}_J^{B\beta} {\cal U}_P^{B\gamma} {\cal U}_Q^{D\delta} 
{\cal W}_{\alpha \beta \gamma \delta} \,.
\ee
The Ricci and scalar curvatures are completely fixed by the constant $\lambda$, independently of the tensor ${\cal W}_{IJPQ}$ 
which turns out to have vanishing contractions, and read ${\cal R}_{IJ} = 2 (n + 2) \lambda\, {\mathpzc g}_{IJ}$ and 
${\cal R}_{\rm sca} = 8 n (n + 2) \lambda$. The Weyl curvature, on the other hand, does depend on the tensor 
${\cal W}_{IJPQ}$. Finally, one can also write a Ricci decomposition of the curvature. Since the space 
is Einstein, this simply reads:
\bea
{\cal R}_{IJPQ} = \frac {4 \lambda (n+2)}{4n-1} {\mathpzc g}_{I[P} {\mathpzc g}_{JQ]} + {\cal C}_{IJPQ} \,.
\label{RiemannN2Ricci}
\eea
Notice that in general the first and second parts of (\ref{RiemannN2}) and  (\ref{RiemannN2Ricci}) cannot be 
separately identified, and the tensor ${\cal W}_{IJPQ}$ does therefore not exactly coincide with the Weyl 
curvature ${\cal C}_{IJPQ}$. The reason for this is that thanks to the existence of the three complex structure one can 
actually construct linear combinations of the first three terms in (\ref{RiemannN2}) which have vanishing trace. 
It is straightforward to verify that the two pieces in (\ref{RiemannN2}) and (\ref{RiemannN2Ricci}) only separately 
coincide in the minimal case $n=1$, which we shall study in some detail later.

Let us next describe the properties of the Killing vector $k^I$ that is used to generate a potential
for the scalars through a gauging with the graviphoton ${\cal A}_\mu$. First of all, $k^I$ has to satisfy the 
usual Killing equation $\nabla_{(I} k_{J)} = 0$. Moreover, it has to be triholomorphic and thus admit 
a triplet of real Killing prepotentials ${\cal P}^x$, such that $\nabla_I {\cal P}^x = 2\lambda\,{\cal J}^x_{IJ} k^J$ and
$k_I = -\frac 1{6\lambda}\, {\cal J}^x_{IJ} \nabla^J {\cal P}^x$.  It is then convenient to introduce the following quantities:
\bea
{\cal P}^x \,,\;\; {\cal N}^I = 2\hspace{1pt} k^I \,.
\eea
These satisfy
\bea
\nabla_I {\cal P}^x = \lambda\hspace{1pt} {\cal J}^x_{IJ}\hspace{1pt} {\cal N}^J \,.
\eea
Moreover:
\bea
\big[\nabla_I, \nabla_J\big] {\cal P}^x = 2\lambda\,\epsilon^{xyz} {\cal J}^y_{IJ} {\cal P}^z \,,\;\;
\big[\nabla_I, \nabla_J\big] {\cal N}_P = {\cal R}_{IJPQ}\hspace{1pt} {\cal N}^Q \,.
\label{commN2}
\eea
The scalar potential then takes the following form:
\bea
{\cal V} = - \lambda \hspace{2pt} {\cal N}^I {\cal N}_I - 3\hspace{1pt} {\cal P}^x {\cal P}^x \,.
\label{potN2}
\eea
Under supersymmetry transformations, the gravitino and the hyperini transform 
as $\delta \psi_\mu^A = \frac 12 {\cal P}^x \sigma^{xAB} \gamma_\mu \epsilon_B + \cdots$ and
$\delta \xi^\alpha = \sqrt{\text{--}\lambda}\, {\cal N}^{A\alpha} \epsilon_A + \cdots$, where 
${\cal N}^{A \alpha} = {\cal U}^{A \alpha}_I {\cal N}^I$. 
The supersymmetry breaking scale is thus linked to the norm of ${\cal N}_I$, $M_{\rm susy} = \sqrt{\text{--} \lambda} \,|{\cal N}|$, 
while the gravitino masses are given by the norm of ${\cal P}^x$, $m_{3/2} = |{\cal P}|$.
The two independent Goldstini are $\xi^A = {\cal N}^A_\alpha \xi^\alpha$, while the corresponding 
four real sGoldstini are $q^{AB} = {\cal N}_I^{AB} q^I$, where 
${\cal N}_I^{AB} = {\cal U}_I^{A \alpha} {\cal N}_\alpha^B = \frac 12 \epsilon^{AB} {\cal N}_I + \frac i2 \sigma^{xAB} ({\cal J}^x {\cal N})_I$.
We then see that the singlet sGoldstino $q = {\cal N}_I q^I$ is actually just the would-be Goldstone mode absorbed
by the graviphoton in the Higgs mechanism giving it mass, while the triplet sGoldstini $q^x = ({\cal J}^x {\cal N})_I q^I$ are instead 
physical modes. 

The mass matrix of the scalars at a stationary point, where $\nabla_I {\cal V} = 0$, is given by 
${\mathpzc m \!}^2_{IJ} = -\frac 1{2\lambda} \nabla_I \nabla_J {\cal V}$. This is not yet the physical mass matrix, 
but having included the factor of $-\frac 1{2\lambda}$ to partly compensate the non-canonical normalization 
of the kinetic terms, the only thing that is left is the effect of the non-trivial metric 
${\mathpzc g}_{IJ}$. The physical masses along the direction ${\mathpzc n\!}^I = {\cal N}^I/|{\cal N}|$ and in the 
subspace of directions ${\mathpzc n\!}^{xI} = ({\cal J}^x {\cal N})^I/|{\cal N}|$ are then simply given by 
${\mathpzc m\!}^2 = {\mathpzc m\!}^2_{IJ} {\mathpzc n\!}^I \hspace{-1pt} {\mathpzc n\!}^J$ and 
${\mathpzc m\!}^{2xy} = {\mathpzc m\!}^2_{IJ} {\mathpzc n\!}^{xI} \hspace{-1pt} {\mathpzc n\!}^{yJ}$. After a straightforward 
computation, one finds that \cite{GRLS}
\bea
{\mathpzc m\!}^2 = 0 \,,\;\; {\mathpzc m\!}^{2xy} = - \big({\cal R}^{xy} - 3\lambda\, \delta^{xy}\big) {\cal N}^I {\cal N}_I
- 4 \big({\cal P}^x {\cal P}^y - \delta^{xy} {\cal P}^z {\cal P}^z\big) \,.
\eea
where
\bea
\hspace{-20pt} && {\cal R}^{xy} = \frac {{\cal R}_{IJPQ} \hspace{1pt} 
{\cal N}^I ({\cal J}^x {\cal N})^J {\cal N}^P ({\cal J}^y {\cal N})^Q}{({\cal N}^R {\cal N}_R)^2} \,.
\label{Rxy}
\eea
This shows that the mass of the singlet sGoldstino vanishes, in agreement with the fact that it 
actually corresponds to the would-be Goldstone mode associated to the gauged isometry, while the 
mass submatrix of the triplet sGoldstini is entirely controlled by the bisectional curvatures ${\cal R}^{xy}$ 
of the scalar manifold in the two planes defined by the conjugate vectors $({\cal N},{\cal J}^x {\cal N})$ and 
$({\cal N},{\cal J}^y {\cal N})$.

Now, using the general form (\ref{RiemannN2}) of the Riemann tensor on a quaternionic
manifold, one finds that the bisectional curvatures (\ref{Rxy}) take the following general form in terms 
of ${\cal W}_{\alpha \beta \gamma \delta}$:
\be
{\cal R}^{xy} = 2\lambda\, \delta^{xy} - \frac {{\cal W}_{\alpha \beta \gamma \delta} 
{\cal N}^{A \alpha} {\cal N}^{B \beta} {\cal N}^{C \gamma} {\cal N}^{D \delta}}
{({\cal N}^{E \epsilon} {\cal N}_{E \epsilon})^2 }\ \sigma^x_{AB} \sigma^y_{CD} 
\label{RSigmaxy} \,.
\ee
A crucial property of the diagonal elements of this matrix, which correspond to the sectional curvatures 
in the three planes defined by the conjugate vectors $({\cal N},{\cal J}^x {\cal N})$, is that their average is completely 
universal and fixed by the Ricci curvature. Indeed, when contracting (\ref{RSigmaxy}) with 
$\frac 13 \delta_{xy}$ the term involving ${\cal W}_{\alpha \beta \gamma \delta}$ drops out, as can 
be seen by using the identity $\sigma^x_{AB} \sigma^x_{CD} = - 2\, \epsilon_{A(C} \epsilon_{BD)}$, 
and one finds:
\be
{\cal R}_{\rm avr} = \frac 13 \delta_{xy} {\cal R}^{xy} = 2\lambda \,.
\ee
As a consequence of this result, the average square mass ${\mathpzc m\!}^2_{\rm avr} = \frac 13 \delta_{xy} {\mathpzc m\!}^{2xy}$ of 
the three non-trivial sGoldstini is simply ${\mathpzc m\!}^2_{\rm avr} = \lambda\hspace{2pt} {\cal N}^I {\cal N}_I + \frac 83 {\cal P}^x {\cal P}^x$, 
or equivalently \cite{GRLS}:
\be
{\mathpzc m\!}^2_{\rm avr} = -\, {\cal V} - \frac 13 m_{3/2}^2 \,.
\label{sumrule}
\ee
This represents by construction an upper bound on the square mass of the lightest scalar. We then deduce 
that it is impossible to have a metastable supersymmetry breaking vacuum with ${\cal V}>0$, no matter which 
quaternionic manifold one chooses and which isometry one gauges, since in such a situation the right-hand 
side of eq.~(\ref{sumrule}) is negative and therefore there must be at least one tachyon.

\section{N=2 to N=1 truncations}\setcounter{equation}{0}

Let us now consider the truncation of a generic N=2 theory with $n$ hypermultiplets to an 
N=1 theory with $n$ chiral multiplets. For the geometry, this means that we start from a 
generic quaternionic manifold and then select a K\"ahler submanifold, \footnote{It has
been shown that the maximal dimension of K\"ahler submanifolds of a quaternionic K\"ahler 
manifold of dimension $n$ is $n/2$ \cite{AM}, therefore we are considering the maximal case 
here. See also \cite{CLST} for a recent discussion on methods to obtain K\"ahler manifolds from 
quaternionic manifolds through quotients.} while for the potential this means that we start from the 
gauging of a triholomorphic isometry and get a restricted superpotential out of it. We will follow the 
general discussion of \cite{ADF1,ADF2}.

Let us first consider the reduction of the geometry. The $4n$ real scalar fields must split into 
$2n$ tangent and $2n$ normal real fields. We can correspondingly decompose the curved 
index as $I \to I_\parallel, I_\perp$. The truncation then acts by setting 
\be
q^{I_\perp}| = 0 \,,
\ee
while $q^{I_\parallel}|$ are real coordinates on the K\"ahler submanifold.
The holonomy must reduce from $SU(2) \times SP(2n)$ to $U(n)$. More precisely, since
$SU(2) \supset U(1)$ and $SP(2n) \supset U(n)$, the $U(1)$ part can arise 
from a linear combination of factors coming from $SU(2)$ and $SP(2n)$, while the $SU(n)$ part can 
only come from $SP(2n)$. We can correspondingly decompose the flat indices as $A \to 1,2$ 
and $\alpha \to a, \bar a$. The truncation then acts by setting 
\be
{\cal U}_{I_\parallel}^{2a}| = ({\cal U}_{I_\parallel}^{1\bar a})^*| = 0 \,,
\ee
while ${\cal U}_{I_\parallel}^{1a}| = ({\cal U}_{I_\parallel}^{2 \bar a})^*|$ are vielbeins for the 
K\"ahler submanifold which are complex but not necessarily compatible with its complex structure.
From the reduction of the torsion-free equations for the vielbeins onto the submanifold, one then 
deduces that one should have:
\bea
\omega_{I_\parallel}^1| = \omega_{I_\parallel}^2| = 0 \,,\;\;
\Delta^a_{{I_\parallel} \bar b}| = 0 \,.
\eea
Moreover, Frobenius' theorem dictates an involution condition, which amounts to requiring that the
curvatures of the connections that have been set to zero are also zero. For the $SU(2)$ part this 
is automatic, but for the $SP(2n)$ part this requires an additional condition, which reads:
\bea
{\cal W}_{abc \bar d} | = 0 \,. 
\eea
In such a situation, the metric automatically splits into a block diagonal form, and the first block 
is identified with the metric of the submanifold, modulo a normalization factor dictated by the different 
normalizations of the kinetic terms in the N=2 and N=1 Lagrangians. Similarly, the reduced vielbein 
splits in tangent and normal components, and the former can be identified with the vielbein on 
the submanifold, again modulo a normalization factor. Furthermore, the first two complex structure 
vanish, ${\cal J}_{I_\parallel J_\parallel}^1| = {\cal J}_{I_\parallel J_\parallel}^2| = 0$, while the third one 
is proportional to the complex structure of the submanifold. More precisely, one finds:
\bea
g_{I_\parallel J_\parallel} = -2\lambda\, {\mathpzc g}_{I_\parallel J_\parallel}| \,,\;\;
e^a_{I_\parallel} = \sqrt{\text{--}2\lambda} \hspace{2pt} {\cal U}^{1a}_{I_\parallel}| \,,\;\;
J_{I_\parallel J_\parallel} = - 2\lambda\hspace{1pt} {\cal J}_{I_\parallel J_\parallel}^3| \,.
\eea
The $U(n)$ connection $\Gamma^a_{I_\parallel\hspace{1pt}b}$ can be identified with a definite 
linear combination of the reduced $SU(2)$ and $SP(2n)$ connections:  
$\Gamma^a_{I_\parallel \hspace{1pt}b} = \omega_{I_\parallel}^3| \delta^a_b + \Delta^a_{I_\parallel\hspace{1pt}b}|$. 
The curvature two-form is then similarly given by 
$R^a_{I_\parallel J_\parallel b} = K_{I_\parallel J_\parallel}^3| \delta^a_b + \Sigma^a_{I_\parallel J_\parallel b}|$.
Finally, the reduced Riemann tensor is proportional to the curvature tensor of the submanifold:
\be
R_{I_\parallel J_\parallel P_\parallel Q_\parallel} = - 2\lambda \hspace{1pt} 
{\cal R}_{I_\parallel J_\parallel P_\parallel Q_\parallel}| \,.
\ee
The curvature on the submanifold can be more explicitly expressed in terms of the symmetric tensor characterizing the 
original quaternionic geometry. More precisely, what matters is the suitably rescaled component with two holomorphic 
and two antiholomorphic tangent-space indices, defined as: 
\be
Y_{a \bar b c \bar d} = - \frac 1{2\lambda} \hspace{1pt} {\cal W}_{a \bar b c \bar d}| \,.
\ee
The curvature two-form then reads $R^{a \bar b}_{I_\parallel J_\parallel} = - \frac i2 J_{I_\parallel J_\parallel} \delta^{a \bar b} 
+ e^{c}{\hspace{-3pt}}_{[I_\parallel \raisebox{7pt}{$$}\hspace{1pt}} e_{J_\parallel]}^{\bar d} 
(\delta^a_{c} \delta^{\bar b}_{\bar d} + 2 \,\hspace{1pt} Y^{a \bar b}{\!}_{c \bar d} \big)$.
The curvature tensor with flat indices correspondingly takes the simple form:
\bea
R_{a \bar b c \bar d} = \delta_{a (\bar b} \, \delta_{c \bar d)} + Y_{a \bar b c \bar d} \,.
\eea
Finally, to write its version with curved indices, it is convenient to switch to complex coordinates that are compatible with the 
complex structure, $I_\parallel \to i, \bar \imath$, in which $e^a_i \neq 0$ but $e^a_{\bar \imath} = 0$. One then finds:
\be
R_{i \bar \jmath p \bar q} = g_{i (\bar \jmath} \, g_{p \bar q)} + Y_{i \bar \jmath p \bar q} \,,
\label{RiemannN1}
\ee
where now:
\be
Y_{i \bar \jmath p \bar q} = e^a_i e^{\bar b}_{\bar \jmath} e^c_p e^{\bar d}_{\bar q}\, Y_{a \bar b c \bar d} \,.
\ee
We see that the Ricci and the scalar curvatures have a universal part plus a contribution from the contractions
$Y_{a \bar b} = - \delta^{c \bar d} Y_{a \bar b c \bar d}$ and $Y_{\rm sca} = \delta^{a \bar b} Y_{a \bar b}$:
$R_{i \bar \jmath} = - \frac 12 (n + 1) g_{i \bar \jmath} + Y_{i \bar \jmath}$ and $R_{\rm sca} = - \frac 12 n (n + 1) + Y_{\rm sca}$. 
The Weyl part of the curvature is instead controlled by the traceless part of the tensor $Y_{a \bar b c \bar d}$.
From this we see that the curvature of the K\"ahler submanifold is a priori arbitrary, since the tensor 
$Y_{a \bar b c \bar d}$ characterizing the curvature of the original quaternionic manifold is also arbitrary 
in principle. In particular, there is a priori no restriction on the value of the sectional curvature along a complex 
direction within the submanifold.

Let us next consider the reduction of the gauging. The graviphoton is of course set to zero by the 
truncation: ${\cal A}_\mu| = 0$. In the situation where the graviphoton is used to gauge an isometry 
on the quaternionic manifold, consistency imposes some restrictions on the Killing vector $k^I$ and 
the related Killing prepotentials ${\cal P}^x$. More precisely, open finds 
that one should have ${\cal P}^3| = 0$, while ${\cal P}^1|$ and ${\cal P}^2|$ can be non-vanishing, 
and similarly that  $k^{I_\parallel}| = 0$ while $k^{I_\perp}|$ can be non-vanishing:
\be
{\cal P}^3| = 0 \,,\;\; k^{I_\parallel}| = 0 \,.
\ee
One can then see that the complex combination ${\cal P}^2| - i\hspace{1pt} {\cal P}^1|$ 
can be identified with the covariantly holomorphic section of the truncated theory: 
\be
L = {\cal P}^2| - i\hspace{1pt} {\cal P}^1| \,.
\ee
The reduced fermionic shifts split into vanishing components ${\cal N}^{1a}| = ({\cal N}^{2 \bar a})^*| = 0$ and 
non-vanishing components ${\cal N}^{2a}| = ({\cal N}^{1\bar a})^*| \neq 0$. The later can be identified with the 
fermionic shifts of the truncated theory, modulo a normalization factor: 
\footnote{Notice that by using the reality condition satisfied by ${\cal U}_I^{A\alpha}$, one
finds that $N^{\bar a} = - \sqrt{\text{--} 2 \lambda} \hspace{2pt} {\cal N}^{1 \bar a}$. As a check, one may verify 
that the second relation in eq.~(\ref{commN2}) correctly reduces to the second relation in eq.~(\ref{commN1})
when restricted to the K\"ahler submanifold.}
\be
\bar N^a = \sqrt{\text{--} 2 \lambda} \hspace{2pt} {\cal N}^{2 a} \,.
\ee
With this identification, one then finds that the reduced potential coincides with the potential of 
the truncated theory:
\be
V = {\cal V}| \,.
\ee
Finally, it turns out that the third derivative of the holomorphic section, which enters in the expression for the mass matrix 
and controls in particular the splitting of sGoldstino masses, can also be more explicitly expressed in terms of the 
symmetric tensor characterizing the original quaternionic geometry. More precisely, what matters in this 
case is the suitably rescaled component with four holomorphic tangent-space indices, defined as: 
\be
Z_{a b c d} = - \frac 1{2\lambda} \hspace{1pt} {\cal W}_{a b c d}| \,.
\ee
Using the fact that ${\cal N}^{A\alpha}$ and $\bar N^a$ are defined in terms of a Killing vector, on which one can then act with 
two covariant derivatives to produce a Riemann tensor, it is straightforward to show that
\be
\nabla_a \nabla_b \nabla_c L = - Z_{abcd} \bar N^c \,.
\label{nabla3L}
\ee

We may now check what happens to the sGoldstino mass matrix of the N=2 theory after the truncation and 
compare this with the general form of the sGoldstino mass matrix for N=1 theories. After a straightforward 
computation, one finds that
\bea
\hspace{-20pt} &&{\cal R}^{xy}| = 2 \lambda \delta^{xy} - \lambda \big(\sigma^{(x}_{11} \sigma^{y)}_{22} + 2\, \sigma^x_{12} \sigma^y_{12} \big) Y
- \frac {\lambda}{2} \sigma^x_{22} \sigma^y_{22} Z - \frac {\lambda}{2} \sigma^x_{11} \sigma^y_{11} \bar Z \,, \\
\hspace{-20pt} &&{\cal P}^x| = \frac i2 \big(\sigma^x_{11} L + \sigma^x_{22} \bar L \big) \,,
\eea
where
\be
Y = - \frac {Y_{a \bar b c \bar d} \bar N^a N^{\bar b} \bar N^c N^{\bar d}}{(\bar N^e N_e)^2} \,,\;\;
Z = - \frac {Z_{a b c d} \bar N^a \bar N^b \bar N^c \bar N^d}{(\bar N^e N_e)^2} \,.
\ee
From this we deduce that the reduced mass matrix ${\mathpzc m\!}^{2xy}|$ is block diagonal, and 
recalling that $|{\cal N}|^2 = - \frac 1{\lambda} |N|^2$ we see that its non-vanishing entries are ($\hat x,\hat y = 1,2$):
\bea
\hspace{-20pt} && {\mathpzc m\!}^{2\hat x \hat y}| = \! \left(\begin{matrix}
\big(\!-\!1 \!+\! Y \!-\hspace{-1pt} {\rm Re} Z\big)|N|^2 \!+\hspace{-1pt} 4\hspace{1pt}({\rm Re}L)^2 
\!\!\!\!\!&\!\!\!\!\! {\rm Im} Z\hspace{1pt} |N|^2 \!+\hspace{-1pt} 4\hspace{1pt} {\rm Re}L\, {\rm Im}L \smallskip\ \\
{\rm Im} Z\hspace{1pt} |N|^2 \!+\hspace{-1pt} 4\hspace{1pt} {\rm Re}L\, {\rm Im}L 
\!\!\!\!\!&\!\!\!\!\! \big(\!-\!1 \!+\! Y \!+\hspace{-1pt} {\rm Re} Z\big)|N|^2 \!+\hspace{-1pt} 4\hspace{1pt}({\rm Im}L)^2\! \\
\end{matrix}\right) \! , \hspace{-10pt} \\[1mm]
\hspace{-20pt} && {\mathpzc m\!}^{233}| = \big(-1\hspace{-1pt}-\hspace{-1pt} 2 \hspace{1pt} Y\big)|N|^2 \!+\hspace{-1pt} 4 \hspace{1pt} |L|^2 \,.
\eea
The three eigenvalues of this mass matrix are easily found to be
\bea
\hspace{-20pt} && {\mathpzc m\!}^2_\pm| = \big(\!-\!1+Y\big) |N|^2 \!+ 2\hspace{1pt} |L|^2 
\pm \big|Z \hspace{1pt} |N|^2 \!- 2 \hspace{1pt}  \bar L^2\big| \,, \\[1mm]
\hspace{-20pt} && {\mathpzc m\!}^{2}_{\rm pro}| = \big(\!-\!1- 2 \hspace{1pt}Y\big) |N|^2 + 4 \hspace{1pt} |L|^2 \,.
\eea
The first two eigenvalues correspond to the two sGoldstini of the N=1 truncated theory: 
$m^2_\pm = {\mathpzc m\!}^2_\pm|$. Indeed, they take the expected form
\be
m^2_\pm = R \hspace{1pt} |N|^2 + 2 \hspace{1pt} |L|^2 \pm \big|\Delta |N|^2 - 2  \hspace{1pt} \bar L^2\big| \,,
\label{mavrN2N1}
\ee
in terms of the holomorphic sectional curvature $R$ and the third covariant derivative of the section $\Delta$ evaluated 
along the supersymmetry breaking direction, which as a consequence of eqs.~(\ref{RiemannN1}) and (\ref{nabla3L}) are 
related to $Y$ and $Z$ as follows:
\be
R = - 1 + Y \,,\;\; \Delta = Z \,.
\label{relYZ}
\ee
The third eigenvalue instead corresponds to the remaining original sGoldstino that is projected out from the 
N=1 theory: $m^2_{\rm pro} = {\mathpzc m\!}^2_{\rm pro}|$. 
Summarizing, the average mass $m^2_{\rm avr} = \frac 12(m^2_+ + m^2_-)$ of the two surviving sGoldstini and 
the mass $m^2_{\rm pro}$ of the projected sGoldstino are then given by the following simple expressions in terms 
of $V$ and $m^2_{3/2}$:
\bea
\hspace{-20pt} && m^2_{\rm avr} = R\, V + \big(3 R + 2\big) m^2_{3/2} \,, \\[1mm]
\hspace{-20pt} && m^2_{\rm pro} = - \big(2R +3\big) V - \big(6 R + 5\big) m^2_{3/2} \,.
\eea
As a check, we can verify that these satisfy the sum rule of N=2 theories:
\be
\frac 23\, m^2_{\rm avr} + \frac 13\, m^2_{\rm pro} = -\, V - \frac 13\, m_{3/2}^2 \,.
\ee
The above analysis provides the link between the metastability condition for N=2 theories and the one for 
N=1 theories, and shows what may happen in a truncation. 
\footnote{A similar analysis was previously performed in \cite{JS} in the rigid limit, where N=2 theories can 
be viewed as particular N=1 theories even without performing any truncation.}
As in the case of general N=1 theories, the quantity $m^2_{\rm avr}$ defines 
by construction an upper bound to the square mass of the lightest scalar, and in order to have a metastable 
supersymmetry breaking vacuum with $V>0$, one needs the sectional curvature $R$ to satisfy 
the bound
\be
R > - \frac {2}{3 + V/m_{3/2}^2} > - \frac 23\,.
\ee
This represents a necessary condition for the existence of metastable de Sitter vacua on the geometry of 
the K\"ahler submanifold. In this case, however, this condition is only necessary and no-longer sufficient. 
This is because it may now no-longer be possible to suitably tune the splitting between the two sGoldstino 
masses, since the superpotential is no-longer arbitrary. More precisely, the quantity $\Delta$ controlling this 
splitting now also has a geometrical expression, and its value may be restricted. In fact, the mass splitting
$\Delta m^2 = \frac 12(m^2_+ - m^2_-)$ between the two surviving sGoldstini is given by the following 
simple expression in terms of $V$, $m^2_{3/2}$, and the phase $\delta$ of the section $L$:
\bea
\Delta m^2 = \big|e^{2 i \delta} \hspace{-1pt} Z \hspace{1pt} V 
+ \big(3\hspace{1pt} e^{2 i \delta} \hspace{-1pt} Z - 2\big) m^2_{3/2} \big| \,.
\eea
In order for this to be smaller that the average mass, so that none of the two sGoldstini is tachyonic, 
the quantity $e^{2 i \delta} Z$ should satisfy the following bound:
\be
\Big|e^{2 i \delta} Z  - \frac {2}{3 + V/m_{3/2}^2} \Big| < R + \frac {2}{3 + V/m_{3/2}^2} \,.
\ee
This represents another necessary condition for the existence of metastable de Sitter vacua, this time 
on the geometry of the space complementary to the K\"ahler submanifold. Indeed, the possible values 
of $Z$ are already constrained by the knowledge of the quantities $Z_{abcd}$, even if one treats $\bar N^a$ 
as an arbitrary vector. Of course, the fact that this vector is actually related to a Killing vector then further 
constrains the result. In other words, some crucial restriction on the ability of tuning the superpotential 
to make the sGoldstino mass splitting sufficiently small already descends from the form of the geometry 
of the original quaternionic manifold, independently of the knowledge of which isometries this may admit. 

\section{Models with one hypermultiplet}\setcounter{equation}{0}

Let us now study the problem in the simplest possible context of N=2 theories with one hypermultiplet and 
a graviphoton gauging. For this we need to consider a generic quaternionic manifold of minimal 
dimension four possessing at least one isometry. Such a theory can then be truncated to an N=1 theory with 
one chiral multiplet, which is again the simplest possible case in this family, by selecting a suitable K\"ahler
submanifold.

\subsection{Przanowski-Tod spaces}

Four-dimensional quaternionic manifolds possessing at least one isometry are well known and 
their most general realization goes under the name of Przanowski-Tod spaces \cite{P,T}. With a suitable 
choice of real local coordinates $q^I= r,u,v,t$ and with overall normalization corresponding to a scalar 
curvature ${\cal R}_{\rm sca} = 24\lambda$, the line element takes the following general form:
\be
d {\mathpzc s}^2 = - \frac 1{4\lambda\hspace{1pt} r^2} \Big(f dr^2 + f e^h (du^2 + dv^2) + f^{-1} (dt + \Theta)^2 \Big) \,.
\ee
This depends on a single function $h$ of the three variables $r$, $u$, $v$, which must satisfy 
the three-dimensional Toda equation:
\be
h_{uu} + h_{vv} + (e^h)_{rr} = 0 \,.
\ee
The function $f$ is then related to the function $h$ by
\be
f = 2 - r h_r \,.
\ee
Finally, the $1$-form $\Theta$ is determined, modulo an irrelevant exact form, by the following equation, whose 
integrability is guaranteed by the Toda equation:
\be
d \Theta = \big(f_u \, dv - f_v \, du\big) \wedge dr + (f e^h)_r \, du \wedge dv \,.
\ee
In this general parametrization, the manifest isometry acts as a constant real shift on the variable $t$.

In appendix A, we give a detailed account of the geometry of these Przanowski-Tod spaces. We describe 
the most general way of parametrizing the vielbeins and explain how one may then compute explicitly
the $SU(2)$ and $SP(2)$ connections as well as their curvatures. We present explicit results in 
two different parametrizations, which allow us to analyze two different kinds of truncations. 

\subsection{Lagrangian and masses}

The N=2 theory with a single hypermultiplet spanning a Przanowski-Tod space has a kinetic term
for the four real fields $q^I = r,u,v,t$ with the following metric:
\bea
{\mathpzc g}_{IJ} = - \frac 1{4\lambda\hspace{1pt} r^2} \left(\begin{matrix}
f \!+\! f^{\text{--}1} \Theta_r^2 & f^{\text{--}1} \Theta_r \Theta_u & f^{\text{--}1} \Theta_r \Theta_v & f^{\text{--}1} \Theta_r \\
f^{\text{--}1} \Theta_u \Theta_r  & f e^h \!+\! f^{\text{--}1} \Theta_u^2 &  f^{\text{--}1} \Theta_u \Theta_v & f^{\text{--}1} \Theta_u \\
f^{\text{--}1} \Theta_v \Theta_r & f^{\text{--}1} \Theta_v \Theta_u & fe^h \!+\! f^{\text{--}1} \Theta_v^2 & f^{\text{--}1} \Theta_ v \\
f^{\text{--}1} \Theta_r & f^{\text{--}1} \Theta_u & f^{\text{--}1} \Theta_v & f^{\text{--}1} \\
\end{matrix}\right) \,,
\label{gPT}
\eea
The inverse of this metric, which will be relevant below, is easily computed and takes 
the following form:
\bea
{\mathpzc g}^{IJ} = - 4 \lambda \hspace{1pt} r^2 \! \left(\begin{matrix}
f^{\text{--}1} \!&\! 0 \!&\! 0 \!&\! \text{--} f^{\text{--}1} \Theta_r \\
0 \!&\! f^{\text{--}1} e^{\text{--} h} \!&\! 0 \!&\! \text{--} f^{\text{--}1}\! e^{\text{--}h} \Theta_u \\
0 \!&\! 0 \!&\! f^{\text{--}1}\! e^{\text{--}h} \!&\! \text{--} f^{\text{--}1}\! e^{\text{--}h} \Theta_ v \\
\text{--} f^{\text{--}1}\! \Theta_r \!&\! \text{--} f^{\text{--}1}\! e^{\text{--}h} \Theta_u \!&\! 
\text{--} f^{\text{--}1}\! e^{\text{--}h} \Theta_v \!&\! f \!+\! f^{\text{--}1} (\Theta_r^2 \!+\! e^{\text{--}h} (\Theta_u^2 \!+\! \Theta_v^2)) \\
\end{matrix}\right) .
\label{invgPT}
\eea
We notice that in order for the kinetic energy to be positive the metric should to be positive definite, and for this 
we need the function $f$ to be positive: $f > 0$.

We now want to generate a potential through the gauging of an isometry with the graviphoton. 
For this, we may use the isometry that universally occurs in Przanowski-Tod spaces. In the 
coordinates that we have used, this amounts to a simple shift of the $t$ coordinates, and the 
corresponding Killing vector has a single constant non-vanishing component with overall 
normalization corresponding to the coupling constant: $k^t = g$. The shift vector is then
given by
\bea
{\cal N}^I = \left(\begin{matrix}
0 \\ 0 \\ 0 \\ 2 g
\end{matrix}\right)\,.
\eea
It is straightforward to verify that the Killing vector $k^I$ is triholomorphic. The form of the corresponding Killing 
prepotentials ${\cal P}^x$ depends on the choice of parametrization, but their norm is always given by
\bea
\sqrt{{\cal P}^x {\cal P}^x} = \frac {g}{2\hspace{1pt} r} \,.
\eea
By gauging this isometry with the graviphoton one obtains a potential energy given by 
(\ref{potN2}), which yields in this case \cite{DSTV}
\bea
{\cal V} = \frac {g^2}{r^2} \bigg(\frac 1f - \frac 34\bigg) \,.
\label{VN2t}
\eea
The supersymmetry breaking scale $M_{\rm susy}=|{\cal N}|$ and the gravitino mass $m_{3/2} = |{\cal P}|$ 
are instead given by 
\bea
m_{3/2}  = \frac g{2\hspace{1pt}r} \,,\;\; M_{\rm susy} = \frac g{f^{1/2}\hspace{1pt} r} \,.
\eea

Let us now study the properties of a generic supersymmetry breaking vacuum in this theory 
and compare them with the general results discussed in the previous sections. We first notice 
that ${\cal V}$ does not depend on $t$, reflecting the fact that this is the would-be Goldstone 
boson associated to the shift symmetry. The dependence of ${\cal V}$ on $r$, $u$, $v$, on the 
other hand, is constrained by the fact that $f$ is related to a function $h$ that has to satisfy the 
Toda equation. 
The stationarity condition ${\cal V}_I = 0$ is trivially satisfied for the field $t$ but yields three 
non-trivial equations that must be satisfied on the vacuum for the fields $r, u,v$:
\bea
f_r = - 2 f^2 r g^{-2} {\cal V} \,,\;\; f_u = 0 \,,\;\; f_v = 0 \,.
\eea
At such a point, the mass matrix ${\mathpzc m\!}^2_{IJ} = - \frac 1{2\lambda} {\cal V}_{IJ}$ is then found to be:
\bea
{\mathpzc m\!}^2_{IJ} = - \frac {g^2}{2\lambda\hspace{1pt} r^2 f^2} \left(\begin{matrix}
\xi \hspace{-1pt} - \hspace{-1pt} f_{rr}& - f_{ru} & - f_{rv} & \; 0\, \\
- f_{ur} & - f_{uu} & - f_{uv} & \; 0\, \\
-f_{vr} & -f_{vu} & - f_{vv}& \; 0\, \\
0 & 0 & 0 & \; 0\, \\
\end{matrix}\right) \,,
\eea
where 
\bea
\xi = - \frac {3 f^3}{2\hspace{1pt}r^2} \big(r^2 g^{-2} {\cal V} - 4\hspace{1pt} r^4 g^{-4} {\cal V}^2 \big) \,.
\eea
We may now compute the quantity ${\mathpzc m\!}^2_{\rm avr}$ as simply $\frac 13$ of the trace of this 
matrix computed with the inverse metric (\ref{invgPT}), that is 
${\mathpzc m\!}^2_{\rm avr} = \frac 13 \hspace{1pt} {\mathpzc g\!}^{IJ} m^2_{IJ}$. One finds:
\be
{\mathpzc m\!}^2_{\rm avr} = \frac {2\hspace{1pt} g^2}{3 f^3} \big(\xi - f_{rr} - e^{-h} f_{uu} - e^{-h} f_{vv} \big) \,.
\label{mavrN=2}
\ee
But using the definition of $f$ and the Toda equation satisfied by $h$, it is straightforward to show that
$f_{rr} + e^{-h} f_{uu} + e^{-h} f_{vv} = - 2\hspace{1pt} h_{rr} + r h_r \big(3\hspace{1pt} h_{rr} + h_r^2\big)$.
Using then the form of the potential in terms of $f$ and thus $h$ as well as the stationarity conditions, one 
easily verifies that this further implies the following result:
\bea
f_{rr} + e^{-h} f_{uu} + e^{-h} f_{vv} = \frac {f^3}{8\, r^2} \big(1 + 48 \hspace{1pt} r^4 g^{-4} {\cal V}^2\big) \,.
\label{relTodaf}
\eea
It then follows that
\bea
{\mathpzc m\!}^2_{\rm avr} = - \frac {g^2}{12\hspace{1pt} r^2} \big(1 + 12\hspace{1pt} r^2 g^{-2} {\cal V} \big) \,.
\eea
Finally, recalling the form of the potential and the gravitino mass, this can be brought into the 
form that has been shown to hold for the average sGoldstino mass in any theory with only 
hypermultiplets and a graviphoton gauging, namely
\be
{\mathpzc m\!}^2_{\rm avr} = -\, {\cal V} - \frac 13\, m_{3/2}^2 \,.
\ee
As already said, this result implies that for any supersymmetry breaking 
de Sitter vacuum at least one of the scalars is tachyonic and makes the vacuum unstable.

\subsection{Truncation of the first kind}

A first possibility for selecting a K\"ahler submanifold of Przanowski-Tod spaces consists in keeping the coordinates 
$q^{I_\parallel} = r, t$ and discarding the coordinates $q^{I_\perp} = u, v$. A parametrization of the original space that is 
suitable to study this kind of truncation is described in appendix A.2. With the help of the explicit results reported there, 
and using the short-hand notation $z = u + i v$, we can study very explicitly under which conditions the truncation is consistent. 
The condition ${\cal U}^{21}|=0$ is automatically satisfied (since $b|=0$). Moreover, since $d \Theta| = 0$ we can 
locally choose that $\Theta| = 0$ in the surviving vielbein (given by $a|$). 
We then see that the conditions $\omega^1| = \omega^2|=0$ are also automatically satisfied, while the condition 
$\Delta^1_{\;\;\bar 1}\big| = 0$ (where the values $1,\bar 1$ of the flat indices correspond to the values $1,2$)
implies that $f_z| = 0$. The involution condition ${\cal W}_{111\bar 1}| = 0$ is then also automatically satisfied. 
The only condition for the truncation to be consistent is thus that:
\bea
f_z| = 0 \,.
\label{condtrunc1}
\eea

One may now compute the relevant quantities characterizing this truncation of the geometry, 
by using the results collected in appendix A.2 and the definitions of section 4. The line element 
$ds^2 = - 2 \lambda\hspace{1pt} d {\mathpzc s}^2 |$ is
\bea
ds^2 = \frac 1{2\hspace{1pt} r^2} \Big(f|\, dr^2 + f^{-1}| dt^2 \Big) \,.
\eea
The $U(1)$ connection $\Gamma^1_{\;\;1} = \omega^3 | + \Delta^1_{\;\;1}|$ reads
\bea
\Gamma^1_{\;\;1} = \frac {2 f^2 \!+ i \big(4 f \!-\! f^2 \!+ 2\hspace{1pt} r f_r\big)}{4\hspace{1pt} r f^2} \bigg|\, dt \,.
\eea
The K\"ahler form $J = - 2 \lambda {\cal J}^3|$ and the two distinct quantities 
$Y_{1\bar 1 1 \bar 1} = - \frac 1{2 \lambda} {\cal W}_{1 \bar 1 1 \bar 1}|$ and 
$Z_{1111} = - \frac 1{2\lambda} {\cal W}_{1111}|$ controlling 
the curvature are finally given by
\bea
\hspace{-20pt}&& J = \frac 1{2 \hspace{1pt} r^2}\, dr \wedge dt \,, \\
\hspace{-20pt}&& Y_{1\bar 1 1\bar 1} = \frac {r^2\big(r f h_{rrr} - f h_r^2 + 2 f_r^2\big)}{f^3} \bigg| \,, \\
\hspace{-20pt}&& Z_{1111} = - \frac {r^2e^{-h} f_{\bar z \bar z}}{f^2} \bigg| \,.
\eea

The kinetic term of the truncated theory is controlled by the metric on the selected submanifold. 
More explicitly, this metric and its inverse are given by the following expressions in terms of the 
real coordinates $q^{I_\parallel} = r,t$:
\bea
g_{I_\parallel J_\parallel} = \frac {1}{2 \hspace{1pt} r^2} \left(\begin{matrix}
f| & 0 \\
0 & f^{-1}| \\
\end{matrix}\right) \,,\;\;
g^{I_\parallel J_\parallel} = 2 \hspace{1pt} r^2 \left(\begin{matrix}
f^{-1}| & 0 \\
0 & f | \\
\end{matrix}\right) \,.
\label{gginvtrunc1}
\eea
It should then be emphasized that the positivity of the kinetic term implies the condition $f| >0$.

To generate a potential, one may now try to gauge the manifest shift symmetry that arises in any Przanowski-Tod 
space. But unfortunately this turns out to be inconsistent with the type of truncation studied in this subsection. 
Indeed, the tangent and orthogonal components of the Killing vector are given by $k^r = 0$, $k^t = g$ and 
$k^u = 0$, $k^v = 0$, and with the adopted parametrization the Killing prepotentials are found to be given by
${\cal P}^1 = 0$, ${\cal P}^2 = 0$ and ${\cal P}^3 = g/(2\hspace{1pt} r)$. We  then see that $k^{I_\parallel}| \neq 0$
and also ${\cal P}^3| \neq 0$, unless $g$ vanishes. In this kind of truncation, the gauging of the shift symmetry is 
therefore impossible. The only way to generate a potential would then be to rely on some additional isometry 
that might arise in specific cases. To describe in full generality such a possibility, one may consider the particular 
subset of Przanowski-Tod spaces which admit a second commuting isometry. The general spaces with this property
are also well known and go under the name of Calderbank-Pedersen spaces \cite{CP}. They can be parametrized
in terms of a simpler potential defined through a function that depends only on two of the coordinates and now 
satisfies a linear partial differential equation. We will however not attempt to describe in full generality the theories that 
one can get in this way. 

\subsection{Truncations of the second kind}

A second possibility for selecting a K\"ahler submanifold of Przanowski-Tod spaces consists in keeping the coordinates 
$q^{I_\parallel} = r, v$ and discarding instead the coordinates $q^{I_\perp} = u, t$. A parametrization of the original space 
that is suitable to analyze this kind of truncation is described in appendix A.3. With the help of the explicit results reported 
there, we can as before study under which conditions the truncation is consistent. 
The condition ${\cal U}^{21}|=0$ (which is $b|=0$) implies that $\Theta|=0$. In order for this to be locally possible, 
one then needs to have $d \Theta | = 0$. Since $d \Theta| = f_u|\, dv \wedge dr$, 
we see that this implies the condition $f_u| = 0$. The conditions $\omega^1| = \omega^2|=0$ 
and $\Delta^1_{\;\;\bar 1}\big| = 0$ (where again the values $1,\bar 1$ of the flat indices correspond to the values $1,2$) 
now require the stronger condition $h_u| = 0$ to hold true. The involution condition ${\cal W}_{111 \bar 1}| = 0$ is then also 
automatically satisfied. 
The only condition for the truncation to be consistent is thus that:
\bea
h_u| = 0 \,.
\label{condtrunc2}
\eea

One may now as before compute the relevant quantities characterizing this truncation of the geometry, 
by using the results collected in appendix A.3 and the definitions of section 4. The line element 
$ds^2 = - 2 \lambda\hspace{1pt} d {\mathpzc s}^2 |$ is
\bea
ds^2 = \frac 1{2\hspace{1pt} r^2} \Big(f|\, dr^2 + f e^h| dv^2 \Big)  \,.
\eea
The $U(1)$ connection $\Gamma^1_{\;\;1} = \omega^3 | + \Delta^1_{\;\;1}|$ reads
\bea
\Gamma^1_{\;\;1} = \frac {e^{-h/2} \big(i f_v\big)}{2\hspace{1pt} f} \bigg|\, dr 
+ \frac {e^{h/2} \big(2f + i \big(f^2 \!-\! f \!- r f_r\big)\big)}{2\hspace{1pt} r f}\bigg|\, dv\,.
\eea
The K\"ahler form $J = - 2 \lambda {\cal J}^3|$ and the two distinct quantities 
$Y_{1\bar 1 1 \bar 1} = - \frac 1{2 \lambda} {\cal W}_{1 \bar 1 1 \bar 1}|$
and $Z_{1111} = - \frac 1{2\lambda} {\cal W}_{1111}|$ controlling 
the curvature are finally given by
\bea
\hspace{-20pt}&& J =  \frac {f e^{h/2}}{2\hspace{1pt}r^2} \bigg|\, dr \wedge dv  \,, \\
\hspace{-20pt}&& Y_{1\bar 1 1\bar 1} = \frac {r^2}{2 f^3} 
\Big[\!-\!\Big(r f \big(2\hspace{1pt} h_{rrr} \hspace{-1pt} +\hspace{-1pt} 3\hspace{1pt} h_r h_{rr} \hspace{-1pt}+\hspace{-1pt} h_r^3\big) 
- f h_r^2 + 2 f_r^2 \Big) \nn \\[-0.5mm]
\hspace{-20pt} && \hspace{67pt} +\, e^{-h} \Big(f \big(2 f_{vv} - f_v h_v\big) - 2 f_v^2 \Big) \Big] \Big| \,, \\
\hspace{-20pt}&& Z_{1111} = 
\frac {r^2}{2 f^3} \Big[\Big(r f \big(2\hspace{1pt} h_{rrr}  \hspace{-1pt}- \hspace{-1pt} 3 h_r h_{rr}  \hspace{-1pt}- \hspace{-1pt} h_r^3\big) 
- 3 f h_r^2 + 6 f_r^2 \Big) \nn \\[-0.5mm]
\hspace{-20pt} && \hspace{67pt} +\, 2\hspace{1pt} i\hspace{1pt} e^{-h/2} \Big(f\big(fh_{rv} + 2 f_{rv}\big) - 3 \big(f h_r + 2 f_r\big) f_v \Big) \nn \\[0mm]
\hspace{-20pt} && \hspace{67pt} +\,e^{-h}\Big(f \big(2 f_{vv}- f_v h_v\big) - 6 f_v^2 \Big)\Big]\Big| \,.
\eea

The kinetic term of the truncated theory is controlled by the metric on the selected submanifold. 
More explicitly, this metric and its inverse are given by the following expressions in terms of the 
real coordinates $q^{I_\parallel} = r,v$:
\bea
g_{I_\parallel J_\parallel} = \frac {1}{2 \hspace{1pt} r^2} \left(\begin{matrix}
f| & 0 \\
0 & f e^h | \\
\end{matrix}\right) \,,\;\;
g^{I_\parallel J_\parallel} = 2 \hspace{1pt} r^2 \left(\begin{matrix}
f^{-1}| & 0 \\
0 & f^{-1} e^{-h} | \\
\end{matrix}\right) \,.
\label{gginvtrunc2}
\eea
Once again, it should be recalled that the positivity of the kinetic term implies the condition $f| > 0$.

To generate a potential, one may again try to gauge of the shift symmetry that arises in any Przanowski-Tod space.
Fortunately, this turns out to be consistent with the type of truncation studied in this subsection. 
Indeed, the tangent and orthogonal components of the Killing vector are given by $k^r = 0$, $k^v = 0$ and 
$k^u=0$, $k^t = g$, and with the adopted parametrization the Killing prepotentials are found to be given by
${\cal P}^1 = g/(2 \hspace{1pt} r)$, ${\cal P}^2 = 0$ and ${\cal P}^3 = 0$. We then see that $k^{I_\parallel}| = 0$
and also ${\cal P}^3| = 0$, independently of the value of $g$. In this kind of truncation, the gauging of the shift 
symmetry is therefore possible, and we shall now study the form of the resulting potential. Of course, one 
may also try to gauge other additional isometries whenever these arise, like for instance in Calderbank-Pedersen 
spaces, but we shall not study this possibility.
Coming back to the gauging of the manifest shift symmetry of $t$ in any Przanowski-Tod space,
we see that this induces the following non-trivial result for the section $L = {\cal P}^2| - i \hspace{1pt} {\cal P}^1|$:
\be
L = - i \frac {g}{2r} \,.
\ee
The corresponding scalar potential of the truncated theory is then a function of the real fields $r$ and $v$, 
which is just the restriction of the N=2 potential (\ref{VN2t}): 
\bea
V = \frac {g^2}{r^2} \bigg(\frac 1f - \frac 34\bigg) \bigg| \,.
\label{Vtrunc}
\eea
The gravitino mass and the supersymmetry breaking scale are instead given by 
\bea
m_{3/2}  = \frac g{2\hspace{1pt}r} \,,\;\; M_{\rm susy} = \frac g{f^{1/2}\hspace{1pt} r} \bigg| \,.
\eea

Let us finally study the properties of the stationary points of the above potential for the truncated theory
and investigate the possibility of getting metastable de Sitter vacua. The stationarity conditions 
$V_{I_\parallel} = 0$ with respect to the two fields $r, v$ imply respectively the 
following conditions on the vacuum:
\bea
f_r| = - 2 f^2 r g^{-2} V \big| \,,\;\; f_v| = 0 \,.
\eea
At such a point, the mass matrix $m^2_{I_\parallel J_\parallel} = V_{I_\parallel J_\parallel}$ 
is then found to be
\bea
m^2_{I_\parallel J_\parallel} = \frac {g^2}{r^2 f^2} \left(\begin{matrix}
\xi \hspace{-1pt} -\hspace{-1pt} f_{rr}& - f_{rv} \\
-f_{vr} & - f_{vv} \\
\end{matrix}\right) \bigg| \,,
\eea
where now
\bea
\xi = - \frac {3 f^3}{2\hspace{1pt}r^2} \big(r^2 g^{-2} V - 4\hspace{1pt} r^4 g^{-4} V^2 \big) \Big| \,.
\eea
To compute the two physical masses, one needs to take into account the non-trivial metric,
which is given by (\ref{gginvtrunc2}). The simplest way to do so is to first raise an index of the 
mass matrix with the inverse metric to compute $m^{2I_\parallel}{}_{\!\!J_\parallel}$, and then 
derive the two physical masses as the eigenvalues of this matrix: 
$m^2_\pm = {\rm eigenvalues} \big(m^{2I_\parallel}{}_{\!\!J_\parallel}\big)$.
By doing so one arrives at the following expression, which can be verified to match the general 
result (\ref{mavrN2N1}) with the relations (\ref{relYZ}) after using the stationarity condition and 
the Toda equation:
\be
m^2_{\pm} = \frac {g^2}{f^3} \Big[\big(\xi \hspace{-1pt} -\hspace{-1pt} f_{rr} - e^{-h} f_{vv} \big) 
\pm \sqrt{\big(\xi - f_{rr} + e^{-h} f_{vv} \big){\raisebox{8pt}{}}^2 + 4\, e^{-h} f_{rv \raisebox{5pt}{}}^2}\, \Big] \Big| \,.
\ee
The necessary and sufficient conditions for these masses $m^2_\pm = m^2_{\rm avr} \pm \Delta m^2$ to be positive are that 
$m^2_{\rm avr} > 0$ and $\Delta m^2 < m^2_{\rm avr}$. This requires that
$(f_{rr} + e^{-h} f_{vv})| < \xi$ and $f_{rv}^2| < \big(\xi - f_{rr}\big) f_{vv}$.
Equivalently, one needs the two diagonal elements of the mass matrix to be positive and the off-diagonal 
element to be smaller than their geometrical meaning absolute value, which means:
\bea
f_{rr}| < \xi \,,\;\; f_{vv}| < 0 \,,\;\; |f_{rv}|| < \sqrt{(f_{rr} - \xi) f_{vv}}\, \big| \,.
\label{cond}
\eea
We also note that the mass $m^{2}_{uu} = - \frac 1{2\lambda} {\cal V}_{uu}|$ that the non-trivially projected 
mode would have in the N=2 theory is given by $m^2_{uu} = \frac 1{2\lambda} g^2r^{-2} f^{-2} f_{uu} |$. 
Taking into account that ${\mathpzc g\!}^{uu} = - 4\lambda \, r^2 f^{-1} e^{-h} |$, the physical mass of this 
mode is found to be:
\be
m^2_{\rm pro} = - \frac {2 g^2}{f^3} e^{-h} f_{uu} \bigg| \,.
\ee
As a check, we notice that the average of all the three masses $m^2_+$, $m^2_-$ and $m^2_{\rm pro}$ is given 
by $\frac 23 m^2_{\rm avr} + \frac 13 m^2_{\rm pro}$ and matches (\ref{mavrN=2}), which was shown to take the 
universal negative value $-\, {\cal V} - \frac 13 m_{3/2}^2$. We then know that 
$m^2_{\rm avr}  = - \frac 32\, V - \frac 12 m_{3/2}^2 - \frac 12 m^2_{\rm pro}$, 
and achieving $m^2_{\rm avr} > 0$ thus requires that $m^2_{\rm pro} < - 3\, V - m_{3/2}^2$. 
We can verify that this bound on the square mass of the projected state indeed follows from the 
conditions (\ref{cond}) ensuring the positivity of the square masses of the retained states, as a 
consequence of the relation (\ref{relTodaf}). Taking the reversed logic, we can say that this bound 
on $m^2_{\rm pro}$ represents an alternative form of the condition for $m^2_{\rm avr}$ to be positive, 
which exploits the fact that $h$ must satisfy the Toda equation. More precisely, this condition 
takes the following form, which shows in a very simple and direct way that metastable Sitter vacua 
are forbidden whenever the function $f$ does not depend on the variable $u$, no matter how clever the 
dependence on $r$ and $v$ is:
\bea
f_{uu}| > \frac {f^3}{8\hspace{1pt} r^2} e^h \big(1 + 12\hspace{1pt} r^2 g^{-2} V \big)\,\bigg| \,.
\label{condbis}
\eea
We finally conclude that in this truncation it is possible for metastable de Sitter vacua to exist, but only for 
suitable choices of $h$ and thus $f$. For example, to achieve the simplest and most interesting situation 
of a metastable vacuum with $V \simeq 0$, one needs to find a point where $f \simeq \frac 43$, then 
$f_r|\simeq0$, $f_v| = 0$, $f_u| = 0$ and finally $f_{rr}| \lsim 0$, $f_{vv}| \lsim 0$, $|f_{rv}|| \lsim \sqrt{f_{rr} f_{vv}}|$, 
$r^2 e^{-h} f_{uu}| \gsim \frac 8{27}$, $f_{ru}|=0$, $f_{vu}| = 0$.

\subsection{Canonical K\"ahler coordinates}

The two types of truncations described in the previous subsections define N=1 theories. These have 
been parametrized by using a pair of real coordinates $q^{I_\parallel}$, but it should in principle
be possible to find a canonical complex coordinate $\phi$, and then deduce a real K\"ahler potential 
$K(\phi,\bar \phi)$ and a holomorphic superpotential $W(\phi)$. More precisely, this complex coordinate 
should be such that the K\"ahler form $J$ and the line element $ds^2$ on the submanifold take the canonical forms
\be
J = i\hspace{1pt} g_{\phi \bar \phi }\, d\phi  \wedge d \bar \phi  \,,\;\; 
ds^2 = 2\hspace{1pt} g_{\phi \bar \phi}\, d\phi \otimes d \bar \phi \,.
\ee
Once this coordinate has been found, one can derive the K\"ahler potential and the superpotential
from the following relations:
\bea
g_{\phi \bar \phi} = \nabla_\phi  \nabla_{\bar \phi } K \,,\;\; L = e^{K/2} W \,.
\eea
In practice, however, finding the explicit coordinate change from the real variables $q^{I_\parallel}$ to 
the canonical complex coordinate $\phi$, and then deriving  the forms of $K$ and $W$, is not so easy. Indeed, this 
involves solving a differential equation, and it does not seem to be possible to find a closed form solution 
for a completely general Przanowski-Tod space parametrized in terms of the real potential $h$. On the other 
hand, there also exists a different parametrization of such spaces where complex coordinates are used from 
the beginning and a different potential satisfying a slightly different differential equation is involved \cite{Pcomplex}. 
Here we will not attempt to discuss any further the general derivation of the canonical coordinate $\phi$ and 
the explicit forms of $K$ and $W$, since as we saw this is not needed to study the physical properties 
of the truncated model. Instead, we shall describe how this works in some simple examples in next section.

\section{Specific examples}\setcounter{equation}{0}

To illustrate the results derived in previous section for a generic N=2 theory with a single hypermultiplet 
truncated to an N=1 theory with a single chiral multiplet, let us now describe a few explicit examples. 
For this purpose, we shall consider some simple classes of Przanowski-Tod spaces, based on explicit 
solutions of the Toda equation for $h$, and study the implementation of the two kinds of truncations that 
we have described in general terms. We will mostly focus on those cases which are relevant for the description 
of the low-energy effective dynamics of the so-called universal hypermultiplet of N=2 superstrings, 
which after truncation maps to the universal dilaton of N=1 superstrings. 

Let us provide some further details on the application to ten-dimensional superstrings compactified on a rigid 
Calabi-Yau manifold, following \cite{DSTV,STV}. We start from the type IIA superstring and focus our attention 
on the universal sector of scalar 
fields that always arise as a consequence of the existence of a K\"ahler two-form and a holomorphic three-form
which are harmonic. This sector contains the dilaton $\phi$ and the axion $\sigma$ originating from the metric 
and the Kalb-Ramond antisymmetric field in the NSNS sector, and the two scalars $\chi$ and $\varphi$ originating 
from the three-index antisymmetric field in the RR sector. We can then identify the Przanowski-Tod coordinates 
$r,u,v,t$ in the following way in terms of these fields: $r = e^\phi$, $u = \chi$, $v=\varphi$, $t = \sigma$. In this way,
the manifest symmetry shifting $t$ in the Przanowski-Tod framework corresponds to a symmetry shifting the 
field $\sigma$, which can indeed be argued to be a good symmetry of the theory under some mild assumptions.
Moreover, the gauging of this symmetry and the associated potential can naturally emerge in the string context
as the result of a non-vanishing flux for the RR three-form over the Calabi-Yau. Let us then see how the two kinds 
of truncations we described in previous section act in this framework. The first truncation we considered retains 
$r$, $t$ and discards $u$, $v$. It therefore corresponds to a kind of heterotic truncation. The second projection 
we considered retains $r$, $v$ and discards $u$, $t$. It therefore corresponds to an orientifold truncation.
This is very similar to what discussed in \cite{DFTV}.

\subsection{Exact metrics depending on a single coordinate}

The simplest class of Przanowski-Tod spaces is defined by a function $h$ that only depends on $r$ and is a 
solution of the simplified Toda equation $(e^h)_{rr} = 0$. The function $f$ then also depends only on $r$, and the 
form $\Theta$ has a simple universal expression. The most general solution of this type can in fact by studied exactly 
and turns out to be parametrized by a single real constant $c$. Restricting the coordinates to $r>\max\{-c,-2c\}$ 
in such a way that the metric is real and positive definite, and keeping an arbitrary real and positive integration 
constant $r_0$ which could be reabsorbed by a rescaling of the coordinates $u$ and $v$, one finds
\be
h = \log \frac {r+c}{r_0} \,,\;\; f = \frac {r+2c}{r+c} \,,\;\; \Theta = \frac 1{2\hspace{1pt} r_0} \big(u\, dv - v\, du \big) \,. 
\ee
In the limit $c \to 0$ this space reduces to the $SU(1,2)/(U(1)\times SU(2))$ coset manifold.
Choosing $r_0 = 1$, this corresponds to taking $h = \log r$, $f = 1$, $\Theta = \frac 12 \big(u\, dv - v\, du \big)$.
In the limit $c \to + \infty$ this space reduces  to the $SO(1,4)/SO(4)$ coset manifold. 
Choosing $r_0 = c$ to simplify things, this corresponds to taking $h = 0$, $f = 2$, $\Theta = 0$.
In the limit $c \to - \infty$ this space remains as a complicated non-symmetric space, 
because the coordinate $r$ cannot be kept finite.

It was show in \cite{CFG,FS,AMTV} that the above space describes the low-energy effective theory of the 
universal hypermultiplet if one ignores non-perturbative effects arising from two-branes and five-branes. 
The constant $c$ parametrizes the perturbative quantum corrections to the classical dynamics, 
and can be vanishing, positive or negative, depending on the topology of the Calabi-Yau manifold. 
The classical moduli space is thus maximally symmetric, while the perturbative corrections lift some 
isometries and leave only four of them. However, it is important 
to emphasize that the above space is truly quaternionic even for finite and large values of $c$. 
\footnote{See also \cite{AADT} for a study of the rigid limit of this corrected moduli space.}

The N=2 theory based on the above space has kinetic terms that are controlled by a metric 
whose inverse has the following non-trivial entries for the fields $r$, $u$, $v$: 
${\mathpzc g\!}^{rr} = - 4 \lambda r^2 (r+c)/(r+2c)$, 
${\mathpzc g\!}^{uu} = - 4 \lambda r^2 r_0/(r+2c)$,  ${\mathpzc g\!}^{vv} = - 4 \lambda r^2 r_0/(r+2c)$.
The gauging of the symmetry shifting $t$ then produces a potential, but this depends only on $r$ 
and not on $u$ and $v$. Its explicit form is:
\be
{\cal V} = \frac {g^2}{r^2} \bigg(\frac {r+c}{r+2c} - \frac 34 \bigg) \,.
\ee
We already argued in full generality that such a potential depending on a single variable 
cannot admit metastable de Sitter vacua. This is because two of the three sGoldstini are 
unavoidably massless and actually correspond to the Goldstone modes of the additional 
spontaneously broken isometries acting as shifts on $u$ and $v$, and the third one must 
then necessarily be tachyonic as a consequence of the sum rule (\ref{sumrule}). 
It is straightforward to verify that this is indeed what happens. 
For $c > 0$, there is an unstable de Sitter stationary point at $r = (1+ \sqrt{5})\, c$ with 
cosmological constant ${\cal V} =  \frac 1{32} (5\sqrt{5}-11)\, g^2 c^{-2}$, gravitino mass 
$m^2_{3/2} = \frac 1{32} (3 - \sqrt{5})\, g^2 c^{-2}$ and physical rescaled masses 
${\mathpzc m\!}^2_r = - \frac 1{16} (7 \sqrt{5} -15) \, g^2 c^{-2}$, ${\mathpzc m\!}^2_u = 0$, 
${\mathpzc m\!}^2_v = 0$, which give as expected ${\mathpzc m\!}^2_{\rm avr} = - {\cal V} - \frac 13 m^2_{3/2}$.
For $c < 0$ there is no stationary point.

Let us now examine the first kind of N=1 truncation discussed in section 5.3. The consistency condition 
(\ref{condtrunc1}) is automatically satisfied. From the expression of the metric in real coordinates, 
or equivalently from the expression (\ref{RiemannN1}) for the Riemann tensor, one computes that 
the sectional curvature is given by $R = - 1 - (1 + 2\hspace{1pt} c/r)^{-3}$. For $c > 0$, this decreases 
from $0$ to $-1$ and finally to $-2$ when $r$ goes from $-c$ to $0$ and then to $+\infty$. For
$c<0$, this increases instead from $-\infty$ to $-2$ when $r$ goes from $-2\hspace{1pt} c$ to $+ \infty$.
The canonical complex coordinate can in this case be taken to be 
$\phi = r + c \log[(r+c)/r_0] + i t$, so that $\partial(\phi + \bar \phi)/\partial r = 2 (r+2c)/(r+c)$. 
One then finds $g_{\phi \bar \phi} = \frac 14\hspace{1pt} r^{-2} (r+c)/(r+2c)$ and 
$K = - \log \big[(k_0/c)\hspace{1pt} r^2/(r+c)\big]$, where $k_0$ is an arbitrary real constant related to the 
ambiguity of K\"ahler transformations. The inverse relation for $r$ in terms of $\phi+\bar \phi$ 
can be obtained in terms of the so-called product-logarithm or Lambert function $P(x)$, which 
is implicitly defined by the relation $P(x) e^{P(x)} = x$. \footnote{This function is usually denoted by 
$W(x)$, but we shall call it here ${\rm P}(x)$ to avoid confusion with the superpotential.}
This function is double-valued for $x \in (-1/e,0)$ and single-valued for $x \in (0,+\infty)$, 
and the two branches are denoted by $P_{0}(x)$ and $P_{-1}(x)$. For our purposes, we will 
however define a single-valued version of $P(x)$ by choosing $P(x) = P_0(x)$ for $x \in (0, +\infty)$ and 
$P(x) = P_{-1}(x)$ for $x \in (-1/e,0)$. In terms of this function $P(x)$, one easily finds that
$r = c \big[P\big(\frac c{r_0}e^{\frac {\phi + \bar \phi + 2c}{2c}}\big)-1\big]$. The K\"ahler potential 
then reads:
\bea
K = - \log \Bigg[k_0 \frac {\big(P\big(\frac c{r_0}e^{\frac {\phi + \bar \phi + 2c}{2c}}\big)-1\big)^2}
{P\big(\frac c{r_0}e^{\frac {\phi + \bar \phi + 2c}{2c}}\big)}\Bigg] \,.
\eea
In the special case of $SU(1,2)/(U(1)\times SU(2))$, which corresponds to the limit $c \to 0$, we choose 
as before $r_0 = 1$ and set for convenience $k_0 = 2\hspace{1pt} c$. In that case we have 
$\phi = r + i t$ and therefore $r = (\phi + \bar \phi)/2$. The sectional curvature is now constant and given by 
$R=-2$. Finally, the K\"ahler potential is simply $K = - \log \big(\phi + \bar \phi\big)$.
In the special case of $SO(1,4)/SO(4)$, which corresponds to the limit $c \to + \infty$, we choose 
as before $r_0 = c$ and set for convenience $k_0 = 16\hspace{1pt} c^2$. In that case we have 
$s = 2 r + i t$ and thus $r = (\phi + \bar \phi)/4$. The sectional curvature is again constant and given 
by $R= - 1$. Finally, the K\"ahler potential is simply $K = - 2 \log \big(\phi + \bar \phi\big)$.
\footnote{These results for the two basic coset quaternionic manifolds were also obtained in \cite{AEVV}.}

For this first kind of N=1 truncation, it is not possible to gauge the symmetry shifting $t$. 
One may on the other hand try to gauge one of the other three isometries possessed by 
the spaces under consideration. We have verified that only two of these isometries can 
be gauged compatibly with this truncation, and that they both lead to a constant superpotential. 
The corresponding scalar potential does however not posses any stationary point.

Let us next examine the second kind of N=1 truncation discussed in section 5.4. The consistency condition 
(\ref{condtrunc2}) is automatically satisfied. From the expression of the metric in real coordinates, 
or equivalently from the expression (\ref{RiemannN1}) for the Riemann tensor, one finds that 
the sectional curvature is $R = - 1 + \frac 12 (1 + 2\hspace{1pt}  c/r)^{-3}$. For $c > 0$, 
this increases from $-\frac 32$ to $-1$ and finally to $- \frac 12$ when $r$ goes from $-c$ to $0$ and 
then to $+\infty$. For $c<0$, this decreases from $+\infty$ to $- \frac 12$ when $r$ goes from 
$-2\hspace{1pt}c$ to $+ \infty$. The canonical complex coordinates can 
in this case be taken to be $\phi = 2 \big(\sqrt{r_0(r+c)} - |c| \big) + i v$, so that 
$\partial(\phi + \bar \phi) = 2 \sqrt{r_0/(r+c)}$. One then finds $g_{\phi \bar \phi} = \frac 14\hspace{1pt} r^{-2} (r+c)/r_0$
and $K = - 2 \log \big[16\hspace{1pt} k_0\hspace{1pt} r_0\hspace{1pt} r \big]$, where 
$k_0$ is an arbitrary real constant related to the choice of K\"ahler frame. 
The inverse relation for $r$ in terms of $\phi+\bar \phi$ is simply given by
$r = \frac 1{16}\hspace{1pt} r_0^{-1}\big[(\phi+\bar \phi + 4\hspace{1pt} |c|)^2 - 16\hspace{1pt} r_0 \hspace{1pt} c\big]$.
The K\"ahler potential then reads:
\bea
K = - 2 \log \Big[k_0 \Big(\big(\phi + \bar \phi + 4\hspace{1pt} |c| \big)^2\! - 16\hspace{1pt} r_0 \hspace{1pt} c \Big)\Big] \,.
\eea
In the special case of $SU(1,2)/(U(1)\times SU(2))$ corresponding to the limit $c \to 0$, we choose as 
before $r_0 = 1$ and take for convenience $k_0 = \frac 14$. In that case we have $\phi = 2 \sqrt{r} + i v$, and thus 
$r = (\phi + \bar \phi)^2/4$. The sectional curvature is now constant and given by $R = -\frac 12$. Finally, 
the K\"ahler potential is just $K = - 4 \log \big(\phi + \bar \phi\big)$.
\footnote{This agrees with the general result derived in \cite{GL}.}\linebreak
In the special case of $SO(1,4)/SO(4)$, which corresponds to the limit $c \to + \infty$, we choose 
as before $r_0 = c$ and take $k_0 = \frac 1{8}\hspace{1pt} c^{-1}$. In that case we have $s = r + i v$ 
and thus  $r = (\phi + \bar \phi)/2$. The sectional curvature is constant and given by $R = -1$. Finally, 
the K\"ahler potential is simply $K = - 2 \log \big(\phi + \bar \phi)$.

For this second kind of N=1 truncation, it is possible to gauge the symmetry shifting $t$. 
The section of the truncated theory then takes the value $L = - ig/(2r)$. 
Recalling the definition of the complex coordinate $\phi$ and the form of the K\"ahler 
potential that we just derived, and working in terms of the real variable $r$, one 
easily finds that $W = - 8\hspace{1pt} i g \hspace{1pt} k_0\hspace{1pt} r_0$. Since 
this is constant, the change to the complex coordinate has no effect and one simply 
finds:
\bea
W = - 8\hspace{1pt} i g \hspace{1pt} k_0\hspace{1pt} r_0 \,.
\eea
In the special case of the space $SU(1,2)/(U(1) \times SU(2))$ corresponding to $c \to 0$, with 
the choices $r_0=1$ and $k_0 = \frac 14$ also used in the study of $K$, one finds $W =  - 2\hspace{1pt} i g$.
In the special case of the space $SO(1,4)/SO(4)$ corresponding to $c \to + \infty$, with $r_0 = c$
and $k_0 = \frac 1{8}\hspace{1pt} c^{-1}$, one finds instead $W = - i g$. Notice finally 
that one may also try to gauge one of the other three isometries possessed by the spaces 
under consideration. We have verified that two of them can be gauged compatibly with the 
truncation, and lead to a linear and a quadratic superpotential. In the first case, the scalar 
potential admits no stationary point, while in the second case, one finds a supersymmetric 
AdS stationary point. Considering the gauging of a linear combination of the admissible 
isometries doesn't seem to give any novelty either. We thus stick to the gauging of the 
symmetry shifting t.

The physics of this simple truncation is very similar to that of the original theory. 
The kinetic terms are controlled by a metric whose inverse has the following non-trivial 
entries for the fields $r$, $v$: $g^{rr} = 2 r^2 (r+c)/(r+2c)$, $g^{vv} = 2 r^2 r_0/(r+2c)$.
The gauging of the symmetry shifting $t$ then produces a potential, but this depends only 
on $r$ and not on $v$. This has the same form as in the original theory, since the discarded 
modes correspond to the would-be Goldstone mode absorbed by the graviphoton and to a flat 
direction:
\bea
V = \frac {g^2}{r^2} \bigg(\frac {r+c}{r+2c} - \frac 34\bigg) \,.
\eea
Consequently, the structure of the stationary points of this potential is exactly the same as in the 
original theory. For $c > 0$, there is an unstable de Sitter stationary point at $r = (1+ \sqrt{5})\, c$ with 
$V =  \frac 1{32} (5\sqrt{5}-11)\, g^2 c^{-2}$, $m^2_{3/2} = \frac 1{32} (3 - \sqrt{5})\, g^2 c^{-2}$,
$R = \frac 12 (\sqrt{5} - 4)$ and physical masses $m^2_r = - \frac 1{16} (7 \sqrt{5} -15) \, g^2 c^{-2}$, $m^2_v = 0$,
which give as expected $m^2_{\rm avr} = R V + (3 R + 2) m^2_{3/2}$.
For $c < 0$ there is again no stationary point.

\subsection{Approximate metrics depending on three coordinates}

A more general class of Przanowski-Tod spaces is defined by a function $h$ that depends not 
only on $r$ but also on $u$ and $v$ and is a solution of the full Toda equation equation 
$h_{uu} + h_{vv} + (e^h)_{rr} = 0$. Unfortunately the solutions of such an equation 
are more difficult to characterize. It has been argued in \cite{DSTV} that one family of such 
solutions corresponds to a deformation of the one discussed in previous section through 
terms which are exponentially suppressed for large $r$. More precisely, the general form of 
this solution has been shown to be 
\be
h = \log \bigg[r+c + \sum_{n=1}^{+\infty} \sum_{m = 0}^{+\infty} 
\kappa_{n,m} (u,v) (r+c)^{1-m/2} e^{- 2n \sqrt{r+c}}\bigg] \,.
\label{solinst}
\ee
However, the exact determination of the form that the functions $\kappa_{n,m} (u,v)$ must take in order
for this to represent a solution of the Toda equation is a difficult problem, on which 
some progress has been recently achieved by relying on twistor techniques (see for example 
\cite{APV} for an overview). On the other hand, it is rather straightforward to find approximate solutions
of the above form, which solve the Toda equation only to some finite degree of accuracy. This 
can be done by performing an expansion for large values of $r$, in such a way that terms with 
higher and higher values of $n$ and $m$ in (\ref{solinst}) are more and more suppressed. 
At leading order in such a large $r$ expansion, one can then take
$h \simeq \log (r) + c/r + \kappa(u,v)\, e^{- 2\sqrt{r}}$. 
This approximately satisfies the Toda equation in the limit of large $r$ provided the function $\kappa$
satisfies the linear equation $\kappa_{uu} + \kappa_{vv} \simeq - \kappa$. We can then take 
$\kappa(u,v) \simeq A \cos (u + \alpha) + B \cos (v + \beta)$, where $A$, $B$, $\alpha$, 
$\beta$ are arbitrary real constants. In this way, we arrive at the following approximate solution:
\be
h \simeq \log \hspace{1pt}(r) + \frac cr + \big(A \cos (u + \alpha) + B \cos (v + \beta) \big)\hspace{1pt} e^{- 2 \sqrt{r}} \,.
\label{solinstapprox}
\ee

In the context of the effective theory describing the universal hypermultiplet of type II string theory, 
the infinite sum of exponential corrections in (\ref{solinst}) corresponds to non-perturbative instanton 
effects induced by Euclidean two-branes wrapping on three-cycles of the Calabi-Yau \cite{BBS}. 
\footnote{For a similar discussion of the effects of five-branes, see \cite{NS51,NS52}.}
The large-$r$ 
expansion corresponds instead to a weak-coupling expansion and the three terms in (\ref{solinstapprox})
correspond respectively to the leading classical contribution, the one-loop quantum correction and the one-instanton
quantum correction. It should however be emphasized that the exact $u$ and $v$ dependence dictated 
by the Toda equation is obtained only by including the infinite series of instanton corrections as in (\ref{solinst}),
while the one-instanton approximation (\ref{solinstapprox}) only provides an approximate solution to 
the Toda equation.

The gauging of the symmetry shifting $t$ produces a potential for the N=2 theory based on the above
approximate space, which now depends not only on $r$ but also on $u$ and $v$. Working in the large 
$r$ approximation and using the expression (\ref{solinstapprox}) for $h$, one deduces that 
$f \simeq 1 + c/r + (A \cos (u + \alpha) + B \cos (v + \beta)) \,\sqrt{r} e^{- 2\sqrt{r}}$ 
and $\Theta = \frac 12 (u dv - v du) + (A \sin (u + \alpha) + B \sin (v + \beta))\, r\, e^{- 2\sqrt{r}}$.
One then finds the following approximate form for the potential \cite{DSTV}:
\be
{\cal V} \simeq \frac {g^2}{r^2} \bigg(\frac 14 - \frac cr 
- \big(A \cos (u + \alpha) + B \cos (v + \beta) \big) \sqrt{r}\, e^{- 2\sqrt{r}} \bigg) \,.
\label{VapproxN2}
\ee
Taken as an exact starting point, this leads to a stationary point with positive Hessian matrix and a value of the 
potential that can be adjusted by tuning the parameters $c$, $A$ and $B$ \cite{DSTV}. This is in apparent
contradiction with the no-go theorem of \cite{GRLS} reviewed in section 3, which forbids metastable de Sitter vacua. 
We notice however that while the fields $u$ and $v$ are stabilized by the one-instanton 
effect on its own, the field $r$ is stabilized thanks to a competition between the 
classical and loop effects and the one-instanton effect. This suggests that at such a 
minimum the one-instanton approximation for the potential is actually not reliable. 
Moreover, from the general result ${\mathpzc m\!}^2_{\rm avr} \simeq - \, {\cal V} - \frac 13 m^2_{3/2}$,
we can infer that in a consistent treatment, where all the corrections are included in such a way to 
have an exact solution of the Toda equation, a tachyon must unavoidably arise. This is in fact true not only in the 
weak coupling regime, but for any value of the coupling. 

The fact that the one-instanton approximation breaks down at the putative stationary point is supported by 
noting that the exponential factor $e^{- 2 \sqrt{r}}$ suppressing non-perturbative higher-instanton corrections 
would be of the same order as the power factor $1/r$ controlling perturbative higher-loop corrections. 
This is contrary to the implicit assumptions of \cite{DSTV} when arguing that eq.~(\ref{solinst}) provides an exact solution to 
the Toda equation.
Another way of detecting this problem is to work analytically with the one-instanton approximation 
to the potential, but to pay attention to discard all the effects that would formally be affected by higher-order 
corrections on the basis of their scaling with $r$. 
Proceeding in this way to study the stationarity condition, one finds that the stationary point 
now appears to be unstable. It arises for $r e^{-2 \sqrt{r}} \simeq \frac 12 \frac 1{A+B}$, $u \simeq - \alpha$ and $v \simeq - \beta$, 
with cosmological constant ${\cal V} \simeq \frac 14 g^2 r^{-2}$,  gravitino mass 
$m_{3/2}^2 \simeq \frac 14 g^2 r^{-2}$, and physical rescaled scalar masses 
${\mathpzc m\!}^2_r \simeq - g^2 r^{-3/2}$, ${\mathpzc m\!}^2_u \simeq \frac A{A+B} g^2 r^{-3/2}$,
${\mathpzc m\!}^2_v \simeq \frac B{A+B} g^2 r^{-3/2}$, which give ${\mathpzc m\!}^2_{\rm avr} \simeq 0$. 
This is now compatible with the general result ${\mathpzc m\!}^2_{\rm avr} = - \, {\cal V} - \frac 13 m^2_{3/2}$, 
since both ${\cal V}$ and $m^2_{3/2}$ are of order $r^{-2}$, which is smaller than the leading order $r^{-3/2}$ 
that we are allowed to keep for square masses in this approximation. 
This analysis points again to the conclusion that the metastable de Sitter vacuum that seems to arise from 
(\ref{VapproxN2}) is actually fake. 

Let us now examine the first kind of N=1 truncation discussed in section 5.3. The consistency condition 
(\ref{condtrunc1}) requires that $\alpha = 0$ and $\beta = 0$. One then finds a corrected K\"ahler potential
and kinetic metric. The gauging of the symmetry shifting $t$ is on the other hand not consistent with the truncation, 
and therefore there cannot be any superpotential and potential from this source.

Let us next consider the second kind of N=1 truncation discussed in section 5.4, following \cite{STV}. The consistency condition 
(\ref{condtrunc2}) requires that $\alpha = 0$. But for simplicity we shall also restrict to $\beta = 0$. 
One then finds a corrected K\"ahler potential and kinetic metric. The gauging of the symmetry shifting $t$ is 
in this case possible, and yields a non-trivial and corrected superpotential. The scalar potential for the truncated 
theory is then simply given by the reduction of (\ref{VapproxN2}) and reads:
\be
V \simeq \frac {g^2}{r^2}\bigg(\frac 14 - \frac cr - B \cos v \sqrt{r}\, e^{- 2\sqrt{r}}\bigg) \,.
\ee
At this point, one gets exactly the same situation as in the N=2 theory. The above potential taken 
as an exact starting point appears to admit a metastable de Sitter vacuum with a cosmological 
constant that can be adjusted by tuning the parameters $c$ and $B$ \cite{STV}. 
However, such a vacuum in fact lies outside the range of validity of the above approximated potential, 
for the same reasons as in the N=2 theory, and its existence can therefore not be trusted. 
In this case, there is no reason why consistently including higher-order corrections should unavoidably 
lead to an instability in the N=1 theory. Nevertheless, it remains true that when one starts from a weak-coupling 
regime in the N=2 theory the tachyon that unavoidably arises in such a theory is preserved by the truncation 
and remains in the N=1 theory. On the other hand, it is conceivable that by starting from a strong-coupling regime 
in the N=2 theory the tachyon is discarded by the truncation and disappears from the N=1 theory. But there appears 
to be no way to have computational control over such a situation. 

\subsection{General lessons}

To conclude this section, let us spell out more clearly what are the reasons behind the difficulty in 
achieving a metastable de Sitter vacuum in a truncation of the universal hypermultiplet sector of 
superstrings. The main issue is that in order to be in a weakly-coupled situation, where quantum 
corrections can be treated through an asymptotic expansion, the space in the neighborhood of the 
vacuum point should be a small deformation of the classical geometry $SU(1,2)/(U(1) \times SU(2))$. 
More concretely, this means that the value of $r$ should be large, 
since it corresponds to the inverse coupling. From an N=1 viewpoint, the problem of finding a 
metastable de Sitter vacuum then maps to first achieving a K\"ahler submanifold with a large enough 
curvature and then also a suitable superpotential from a gauging, compatible with the weak-coupling regime. 
Concerning the K\"ahler submanifold, we see it represents an obstruction for the first kind of truncation 
but not for the second. Indeed, starting from the $SU(1,2)/(U(1) \times SU(2))$ quaternionic manifold 
with scalar curvature $24 \lambda$ one gets a different $SU(1,1)/U(1)$ K\"ahler submanifold in the two truncations, 
with curvature given respectively by $-2$ and $-\frac 12$. In situations where the space is deformed but
by a small amount, one will then get submanifolds which are correspondingly slightly deformed 
and whose curvature can therefore not depart substantially from the values $-2$ and $-\frac 12$.
Comparing now with the lower bound $- \frac 23$ for the existence of metastable de Sitter vacua, 
we see that this is badly violated in the first truncation but satisfied in the second truncation. 
Concerning the superpotential, we saw that the gauging of the 
symmetry shifting $t$, which is the only symmetry that can always be gauged and has a clear 
interpretation within the string context, does not do the job. No metastable de Sitter vacuum 
can therefore arise in theses truncations of the universal hypermultiplet in a weak-coupling regime.
On the other hand, as already mentioned, it is not excluded that such a vacuum might arise in 
a strongly coupled regime.

\section{Conclusions}\setcounter{equation}{0}

In this paper, we characterized the conditions under which a metastable de Sitter vacuum may arise 
in a generic N=2 to N=1 supergravity truncation and compared these with the known situations 
of N=2 and N=1 theories, focusing for simplicity on models involving only scalar matter multiplets.
In N=2 theories with hypermultiplets based on a quaternionic manifold with a triholomorphic 
isometry gauged by the graviphoton, metastable de Sitter vacua are excluded due to a sum rule 
satisfied by the triholomorphic sectional curvatures, independently of the gauged isometry \cite{GRLS}. 
In N=1 theories with chiral multiplets based on a K\"ahler manifold and a holomorphic superpotential, 
metastable de Sitter vacua are instead permitted if the sectional curvature of the manifold can be 
sufficiently large and the superpotential can be suitably adjusted \cite{GRS1}. In truncations of N=2 
to N=1 theories, the possibility of achieving a viable vacuum then requires to start 
with a de Sitter vacuum of the mother theory that is unstable but leads to a small enough number 
of tachyonic scalars, and then to arrange that the projection enforcing the truncation eliminates 
all of these tachyonic scalars in such a way as to leave a metastable de Sitter vacuum for the daughter 
theory. 

The general methodology on which our analysis was based is a systematic study of the form of the 
mass submatrix for the sGoldstini scalars, which represent the most severe danger for metastability. 
We first studied the problem in full generality, in terms of the geometry of the initial quaternionic manifold 
and of the K\"ahler submanifold selected by the truncation, as well as the isometries.
We then performed a detailed study of the simplest case of theories with a single hypermultiplet 
truncated to theories with a single chiral multiplet, and described two distinct general ways of 
performing such a truncation. Using the general parametrization of such 
models in terms of Przanowski-Tod spaces based on a Toda potential $h$, we then derived more 
explicitly the conditions that the latter function has to satisfy in order for a metastable de Sitter 
vacuum to emerge after the truncation. Finally, we illustrated these results with a few explicit 
examples, which describe to different levels of accuracy the low-energy effective theory of the 
universal hypermultiplet truncated to a dilaton chiral multiplet in superstring models. This 
allowed us to argue that in such a context no metastable de Sitter vacuum can emerge even 
after a truncation, if one assumes a weakly coupled regime.

An other interesting application of the results that we have obtained would be to study 
models with $n+1$ hypermultiplets based on a so-called dual-quaternionic manifold.
These manifolds naturally emerge in the context of type II superstrings compactified on a Calabi-Yau 
manifold, where they describe the classical moduli space for the universal plus the $n$ additional 
hypermultiplets occurring when the Calabi-Yau space possesses not only a K\"ahler-form but 
$n$ additional harmonic two-forms  \cite{CFG,FS}. They have a geometry that is entirely specified 
by a holomorphic prepotential, and possess the property of always admitting two distinct 
submanifolds of complex dimension $n$, which are respectively special-K\"ahler and only K\"ahler. 
Moreover, it is always automatically possible to consistently truncate the original N=2 
theory based on the dual-quaternionic space to N=1 theories based on either of these 
special-K\"ahler and K\"ahler submanifolds \cite{DST}. These two distinct truncations correspond 
from the ten-dimensional point of view to heterotic and orientifold reductions, and 
represent in some sense the generalization of the two truncations we considered in our study 
of the universal hypermultiplet. It would then be interesting to study the possibility of achieving a 
metastable de Sitter vacuum in such truncated theories by relying on a source of potential that consists 
of a gauging that is compatible with the original N=2 supersymmetry, rather than a generic 
superpotential that is compatible only with the final N=1 supersymmetry. The first situation is 
clearly more restrictive and necessarily emerges at the classical level, where a potential can 
only be inherited from the mother theory through effects like background fluxes. The second 
situation is instead more flexible and may emerge only at the quantum level, where a potential 
can also be induced within the daughter theory through effects like instantons or fermion condensates.
The conditions for the possible existence of metastable de Sitter vacua can be studied 
from an N=1 viewpoint in the second case, along the lines of \cite{CGRGLPS1,FSscal}, 
but they should instead be studied from an N=2 viewpoint in the first case, and it would be 
interesting to know whether this leaves any viable option or not. 

\vskip 20pt
\noindent
{\Large \bf Acknowledgements}
\vskip 10pt
\noindent
We are grateful to S.~Vandoren for useful comments and discussions.
The research of C.~S and P.~S. is supported by the Swiss National Science Foundation under the grant
PP00P2-135164. The research of F.~C. is supported by the Fondazione Ing. Aldo Gini and the European 
Commission under the contract PITN-GA-2009-237920 UNILHC.

\appendix

\section{Geometry of Przanowski-Tod spaces}\setcounter{equation}{0}

In this appendix, we give a detailed account of the geometry of Przanowski-Tod spaces. 
We first explain how to set up the problem and parametrize the vielbein. We then consider 
two different kinds of parametrizations, which are suited to analyze the two kinds of truncations 
considered in the paper, and compute for each of these the connections and the curvatures.

\subsection{General parametrization}

To set up the problem of computing the connection and the curvature, it is convenient
to use the language of differential forms. The basic ingredients are the vielbein $1$-forms 
${\cal U}^{A \alpha}$, which satisfy $({\cal U}^{A \alpha})^*= {\cal U}_{A \alpha}$. 
The holonomy is in this case $SU(2) \times SP(2)$, and both indices $A$ and $\alpha$ run 
over two values. The line element can be rewritten as 
\be
d{\mathpzc s}^2 = {\cal U}^{A \alpha} \otimes {\cal U}_{A \alpha} \,.
\ee
The three hyper-K\"ahler $2$-forms are instead given by
\be
{\cal J}^x = - \frac i2 \sigma^{x\; B}_{\;A} {\cal U}^{A \alpha} \wedge {\cal U}_{B \alpha} \,.
\ee
The $SU(2)$ and $SP(2)$ connections $\omega^A_{\;\;B} = \frac i2\, \sigma^{xA}_{\;\;\;\;B}\, \omega^x$ 
and $\Delta^\alpha_{\;\;\beta}$ are then determined by the torsion-free constraints:
\be
d {\cal U}^{A \alpha} + \omega^A_{\;\;B}\, {\cal U}^{B \alpha} + \Delta^\alpha_{\;\;\beta}\, {\cal U}^{A \beta} = 0 \,. 
\label{torsion-free}
\ee
Finally, the curvatures of these connections take the general form:
\bea
\hspace{-20pt}&& K^x = d \omega^x + \frac 12 \epsilon^{xyz} \omega^y \wedge \omega^z 
= - i\lambda \, \sigma^{x\; B}_{\;A} {\cal U}^{A \alpha} \wedge {\cal U}_{B \alpha} = 2 \lambda {\cal J}^x \,, 
\label{curvSU2} \\
\hspace{-20pt}&& \Sigma^\alpha_{\;\;\beta} = d \Delta^\alpha_{\;\;\beta} + \Delta^\alpha_{\;\;\gamma} \wedge \Delta^\gamma_{\bar \beta}
= \lambda\, {\cal U}^{A\alpha} \wedge {\cal U}_{A \beta} 
+ \frac 12 {\cal W}^\alpha_{\;\;\beta \gamma \delta} {\cal U}^{A \gamma} \wedge {\cal U}_A^\delta \,.
\label{curvSP2}
\eea

In the case under examination, the vielbein is a two-by-two matrix which can be parametrized
in terms of only two independent complex elements $a$ and $b$. This parametrization corresponds 
to the most general way of recasting the structure group of the tangent space from the $SO(4)$ form, 
that is natural when viewing the space as a Riemannian manifold, to the $SU(2) \times SP(2)$ form 
that is natural when viewing the space as a quaternionic manifold. More precisely, one can write:
\be
{\cal U}^{A \alpha} = 
\left(\begin{matrix} 
\bar a \!&\! - \bar b \\
b \!&\! a
\end{matrix}\right) \,.
\ee
With this parameterization, the line element reads
\be
d {\mathpzc s}^2 = a \otimes \bar a + b \otimes \bar b + {\rm c.c.} \,,
\ee
and the three complex structures are given by
\bea
\hspace{-20pt}&& {\cal J}^1 = i \big(a \wedge b - \bar a \wedge \bar b \big) \,,\\
\hspace{-20pt}&& {\cal J}^2 = - \big(a \wedge b + \bar a \wedge \bar b\big) \,, \\
\hspace{-20pt}&& {\cal J}^3 = i \big(a \wedge \bar a + b \wedge \bar b \big) \,.
\eea
One can then work out the form of the connections by solving the torsion-free equations (\ref{torsion-free})
and compute their curvatures by using the definitions (\ref{curvSU2}), (\ref{curvSP2}). To do so, one 
however needs to make a definite choice for $a$ and $b$, because their exterior derivatives get involved.
There are infinitely many possible choices for $a$ and $b$, based on different complex combinations of 
the basic differentials $dq^I=dr,du,dv,dt$. In view of discussing a truncation, however, it will be 
convenient to choose to include in $a$ only the differentials of the preserved coordinates $d q^{i_\parallel}$
and in $b$ only the differentials of the truncated coordinates $dq^{I_\perp}$. We shall now discuss two 
different kinds of parametrization, which are well suited to study the two truncations analyzed in the paper.

\subsection{Explicit parametrization of the first kind}

A first possible parametrization, which is well suited to describe truncations where the coordinates 
$q^{I_\parallel} = r, t$ are kept while the coordinates $q^{I_\perp} = u, v$ are discarded, is based on taking 
\bea
\hspace{-20pt}&& a = \frac 1{\sqrt{\text{--}8\lambda}\, r} \big(f^{1/2} dr + i\, f^{-1/2} (dt + \Theta) \big) \,, \\
\hspace{-20pt}&& b = \frac 1{\sqrt{\text{--}8\lambda}\, r} (f e^h)^{1/2} \big(du + i\,dv \big) \,,
\eea
The $SU(2)$ and $SP(2)$ connections are determined by solving the torsion-free constraint 
in the basis of independent forms $a$, $\bar a$, $b$, $\bar b$. Denoting for short $z= u + i v$, 
one finds
\bea
\hspace{-20pt}&& \omega^1 = - \frac {e^{h/2}}r  dv \,,\;\;
\omega^2 = - \frac {e^{h/2}}r  du \,,\;\; 
\omega^3 = \frac 1{2 r} (dt + \Theta) + \frac {h_v}2 du - \frac {h_u}2 dv  \,,
\eea
and ($\Delta^2_{\;\;2} = - \Delta^1_{\;\;1}$, $\Delta^2_{\;\;1} = - \bar \Delta^1_{\;\;2}$):
\bea
\hspace{-20pt}&& \Delta^1_{\;\;1} = i \frac {4f \!-\! f^2 \!+\! 2 r f_r}{4\, r f^2} (dt + \Theta) 
- i \frac {2 f_v \!+\! f h_v }{4 f} du + i \frac {2 f_u \!+\! f h_u}{4 f} dv \,, \\
\hspace{-20pt}&& \Delta^1_{\;\;2} = 
- \frac {e^{-h/2} f_z}{2 f^2} \big(f dr + i (dt + \Theta)\big) 
+ \frac {e^{h/2}(f\!-\! f^2 \!+\! r f_r)}{2\,r f}\big(du + i dv \big) \,.
\eea
The hyper-K\"ahler forms controlling the $SU(2)$ curvature are then found to be
\bea
\hspace{-20pt} && {\cal J}^1 = - \frac {e^{h/2} }{4\lambda\hspace{1pt} r^2} du \wedge (dt + \Theta) 
+ \frac {f e^{h/2}}{4\lambda\hspace{1pt}r^2} dr \wedge dv \,, \\
\hspace{-20pt} && {\cal J}^2 = \frac {e^{h/2} }{4\lambda\hspace{1pt}r^2} dv \wedge (dt + \Theta) 
+ \frac {f e^{h/2}}{4\lambda\hspace{1pt}r^2} dr \wedge du \,, \\
\hspace{-20pt} && {\cal J}^3 = - \frac 1{4\lambda\hspace{1pt}r^2} dr \wedge (dt + \Theta) 
- \frac {f e^{h}}{4\lambda\hspace{1pt}r^2} du \wedge dv \,,
\eea 
while for the symmetric tensor with flat indices controlling the $SP(2)$ curvature one finds
(${\cal W}_{2222} = \overline {\cal W}_{1111}, {\cal W}_{1222} = - \overline {\cal W}_{1112}$):
\bea
\hspace{-20pt}&& {\cal W}_{1111} = 
\frac {2\lambda\hspace{1pt} r^2e^{-h}}{f^3}\Big[f (f_{\bar z \bar z} - f_{\bar z} h_{\bar z}) - 3 f_{\bar z}^2\Big] \,, \\
\hspace{-20pt}&& {\cal W}_{1112} = 
\frac {\lambda\hspace{1pt} r^2e^{-h/2}}{f^3} \Big[f(f h_{r \bar z} + 2 f_{r \bar z}) -3 (f h_r + 2 f_r) f_{\bar z}  \Big] \,, \\
\hspace{-20pt}&& {\cal W}_{1122} = 
- \frac {2\lambda\hspace{1pt} r^2}{f^3} \Big[f\big(r h_{rrr} - h_r^2 \big) + 2 f_r^2 - e^{-h} f_z f_{\bar z} \Big] \,.
\eea

\subsection{Explicit parametrization of the second kind}

A second possible parametrization, which is well suited to describe truncations where the coordinates 
$q^{I_\parallel} = r, v$ are kept while the coordinates $q^{I_\perp} = u, t$ are discarded, can be obtained 
from the previous one through combined $SU(2)$ and $SP(2)$ transformations both corresponding to 
a $\pi/2$ rotation of $SO(3)$ around the second axis, and is based on taking 
\bea
\hspace{-20pt}&& a = \frac 1{\sqrt{\text{--}8\lambda}\, r} \big(f^{1/2} dr + i\, (f e^{h})^{1/2}dv \big) \,, \\
\hspace{-20pt}&& b = \frac 1{\sqrt{\text{--}8\lambda}\, r} \big((f e^h)^{1/2} du - i\,f^{-1/2} (dt + \Theta) \big) \,.
\eea
The $SU(2)$ and $SP(2)$ connections are again determined by solving the torsion-free 
constraint in the basis of independent forms $a$, $\bar a$, $b$, $\bar b$. One finds:
\bea
\hspace{-20pt} && \omega^1 = - \frac {h_u}2 dv + \frac {h_v}2 du + \frac 1{2 r} (dt + \Theta) \,,\;\;
\omega^2 = - \frac {e^{h/2}}r  du \,,\;\;
\omega^3 = \frac {e^{h/2}}r dv \,,
\eea
and ($\Delta^2_{\;\;2} = - \Delta^1_{\;\;1}$, $\Delta^2_{\;\;1} = - \bar \Delta^1_{\;\;2}$):
\bea
\hspace{-20pt} && \Delta^1_{\;\;1} = i \frac {e^{-h/2} f_v}{2 f} dr 
+ i\frac {e^{h/2} (f^2 \!-\! f \!-\! r f_r)}{2\, r f} dv  + i \frac {e^{-h/2} f_u}{2 f^2} (dt + \Theta)  \,, \\
\hspace{-20pt} && \Delta^1_{\;\;2} = 
- \frac {e^{-h/2} f_u}{2 f} dr + i \frac {f h_u \!+\! 2 f_u}{4 f} dv 
- \frac {2\,e^{h/2}(f^2 \!-\! f \!-\! r f_r) + i r (2 f_v \!+\! f h_v)}{4\, r f} du \nn \\
\hspace{-20pt} && \hspace{35pt} 
-\, i \frac {f^2\!-\! 4f \!-\! 2 r f_r + 2 i r e^{-h/2} f_v}{4\, r f^2} (dt + \Theta) \,.
\eea
The hyper-K\"ahler forms read
\bea
\hspace{-20pt} && {\cal J}^1 = - \frac 1{4\lambda \hspace{1pt} r^2} dr \wedge (dt + \Theta) - \frac {f e^{h}}{4\lambda\hspace{1pt} r^2} du \wedge dv \,, \\
\hspace{-20pt} && {\cal J}^2 = \frac {e^{h/2}}{4\lambda \hspace{1pt} r^2}  dv \wedge (dt + \Theta) 
+ \frac {f e^{h/2}}{4\lambda\hspace{1pt}r^2} dr \wedge du \,, \\
\hspace{-20pt} && {\cal J}^3 = \frac {e^{h/2}}{4\lambda \hspace{1pt} r^2} du \wedge (dt + \Theta) 
- \frac {f e^{h/2}}{4\lambda \hspace{1pt}r^2} dr \wedge dv \,,
\eea
while for the symmetric tensor with flat indices controlling the $SP(2)$ curvature one finds
(${\cal W}_{2222} = \overline {\cal W}_{1111}, {\cal W}_{1222} = - \overline {\cal W}_{1112}$):
\bea
\hspace{-20pt} && {\cal W}_{1111} = 
- \frac {\lambda \hspace{1pt} r^2}{f^3} \Big[3\big(f (r h_{rrr} - h_r^2) + 2 f_r^2 \big) \nn \\[-0.5mm]
\hspace{-20pt} && \hspace{80pt} +\, 2 i e^{-h/2} \big(f(fh_{rv} + 2 f_{rv}) - 3 (f h_r + 2 f_r) f_v \big) \nn \\[0.5mm]
\hspace{-20pt} && \hspace{80pt} +\,e^{-h}\big(f (f_u h_u - f_{uu} - f_v h_v + f_{vv}) - 6 f_v^2 \big)\Big] \,, \\
\hspace{-20pt} && {\cal W}_{1112} = 
- \frac {\lambda \hspace{1pt} r^2}{f^3}\Big[e^{-h/2} \big(f(f h_{ru} +2 f_{ru}) - 3 (f h_r + 2 f_r)f_u  \big) \nn \\[-1mm]
\hspace{-20pt} && \hspace{80pt} +\, i e^{-h} \big(f (h_u f_v + f_u h_v - 2 f_{uv}) + 6 f_u f_v  \big) \Big] \,, \\
\hspace{-20pt} && {\cal W}_{1122} = - \frac {\lambda \hspace{1pt} r^2}{f^3} \Big[\!-\!\big(f (r h_{rrr} - h_r^2) + 2 f_r^2 \big) \nn \\[-1mm]
\hspace{-20pt} && \hspace{80pt} +\, e^{-h} \big(f (f_u h_u - f_{uu} - f_v h_v + f_{vv}) + 4 f_u^2 - 2 f_v^2 \big) \Big] \,.
\eea

\end{document}